\documentstyle[12pt,aaspp4,epsf]{article}
\def\fun#1#2{\lower3.6pt\vbox{\baselineskip0pt\lineskip.9pt
  \ialign{$\mathsurround=0pt#1\hfil##\hfil$\crcr#2\crcr\sim\crcr}}}
\def\lap{\mathrel{\mathpalette\fun <}}
\def\gap{\mathrel{\mathpalette\fun >}}

\def\kms{km s$^{-1}$}

\def\mass{{\cal M}}
\def\Lsolar{{L_\odot}}
\def\Msolar{{\mass_\odot}}

\def\beq{\begin{equation}}
\def\eeq{\end{equation}}

\lefthead{ }
\righthead{Dynamics of M32}

\begin{document}

\title{The Nuclear Dynamics of M32. I.}
\title{ Data and Stellar Kinematics
\footnote{Based on observations with the NASA/ESA {\it Hubble Space Telescope}, obtained at the Space Telescope Science Institute, which is operated by the Association of Universities for Research in Astronomy, Inc. (AURA), under NASA contract NAS5-26555.}}

\bigskip\bigskip

\author{C. L. Joseph,$^2$ D. Merritt$^2$, R. Olling
\footnote{Department of Physics and Astronomy, Rutgers University,
New Brunswick, NJ 08855},
M. Valluri$^{2,}$
\footnote{Astronomy and Astrophysics, University of Chicago, 5640 S. Ellis Avenue, Chicago, IL 60637}, R. Bender,
\footnote{Universit\"ats-Sternwarte, Scheinerstrasse 1, M\"unchen, 81679 Germany}}

\author{G. Bower
\footnote{NOAO, PO Box 26732, Tucson, AZ 85726},
R. F. Green$^5$,
A. Danks
\footnote{RPSC/Goddard Space Flight Center, Code 683, Greenbelt, MD 20771},
T. Gull
\footnote{NASA/Goddard Space Flight Center, Code 681, Greenbelt, MD 20771},
J. Hutchings
\footnote{Dominion Astrophysical Observatory, 5071 W. Saanich Road, Victoria, BC V8X 4M6, Canada},
M. E. Kaiser,
\footnote{Johns Hopkins University, Dept. of Physics \& Astronomy, 34th \& Charles St., Baltimore, MD 21218}}
\author{S. Maran$^7$,
D. Weistrop
\footnote{University of Nevada, Dept. of Physics, 4505 S. Maryland Pkwy., Las Vegas, NV 89154},
B. Woodgate$^7$,
E. Malumuth$^7$,
C. Nelson,$^{10}$}
\author{P. Plait$^7$,
D. Lindler
\footnote{Advanced Comp. Concepts, Inc., 11518 Gainsborough Rd., Potomac, MD 20854}}

\clearpage

\begin{abstract}
We have obtained optical long-slit spectroscopy of the nucleus of M32 using the Space Telescope Imaging Spectrograph aboard the {\it Hubble Space Telescope}.
The stellar rotation velocity and velocity dispersion, 
as well as the full line-of-sight velocity distribution (LOSVD), 
were determined as a function of position along the slit using two 
independent spectral deconvolution algorithms.
We see three clear kinematical signatures of the nuclear black hole:
a sudden upturn, at $\sim 0.3''$ from the center, in the stellar velocity dispersions;
a flat or rising rotation curve into the center; 
and strong, non-Gaussian wings on the central LOSVD.
The central velocity dispersion is $\sim 130$ \kms\ (Gaussian fit) or 
$\gap 175$ \kms\ (corrected for the wings).
Both the velocity dispersion spike and the shape of the central LOSVD are consistent with the presence of a supermassive compact object in M32
with a mass in the range $2-5\times 10^6\Msolar$.
These data are a significant improvement on previous stellar kinematical data, making M32 the first galaxy for which the imprint of the black hole's gravitation on the stellar velocities has been observed with a resolution comparable to that of gas-dynamical studies.

\end{abstract}


\clearpage

\section {Introduction}

The presence of a supermassive compact object, presumably a black hole, 
at the center of the dwarf elliptical galaxy M32 has been suspected for some time (\cite{ton87}).
The evidence consists of rapid rotation of the stars near the center of M32 
and a central peak in the stellar velocity dispersions
(\cite{ton87}; \cite{drr88}; \cite{caj93}; \cite{vdm94a}; \cite{bkd96}).
The most recent study (\cite{vdm97}, 1998) used data from the Faint Object Spectrograph (FOS) on the {\it Hubble Space Telescope} (HST) to infer the rotation and dispersion velocities with a spatial resolution of $\sim 0.1''$ in the inner $\sim 0.5''$ of M32.
The FOS data revealed a sharper rise in the stellar velocity dispersions than had been observed from the ground;
however the velocity resolution of the FOS is limited, making that instrument only marginally useful for the study of a low velocity dispersion system like M32. 
In fact, van der Marel et al. (1998) found large point-to-point variations in their velocity dispersion measurements, making the dynamical interpretation uncertain.
Nevertheless, the case for a supermassive black hole in M32, 
of mass $M_h\approx 3\times 10^6\Msolar$, was considerably strengthened.

Here we present observations of M32 carried out using STIS, 
the Space Telescope Imaging Spectrograph, on HST.
Our data were obtained as part of the STIS Investigation Definition Team's (IDT) key program to observe the nuclei of a sample of $\sim 15$ nearby galaxies in the spectral region centered on the Calcium triplet, $\lambda\approx 8600$\AA.
This paper is the first in a series that will present stellar-kinematical evidence from STIS for the presence (or absence) of supermassive black holes in galactic nuclei.

The STIS data improve on earlier ground-based and FOS data from M32 in several ways.
The spatial resolution of STIS is $\sim 0.1''$, or $\sim 0.3$ pc at the distance of M32, similar to that of the single FOS aperture;
however STIS provides continuous spatial sampling along a slit.
The spectral resolution of STIS in the G750M mode ($\sim 38$ km/s) is much greater than that of the FOS making STIS a more suitable instrument for observing M32, whose velocity dispersion outside of the nucleus is only $\sim 60$ \kms. 
We were able to obtain from the STIS spectra not only the lowest-order moments of the stellar velocity distribution -- the rotation velocity and velocity dispersion -- but also the full line-of-sight velocity distribution (LOSVD) as a function of position along the major axis.

A number of modelling studies (\cite{vdm94b}; \cite{deh95}; \cite{qia95})
have made predictions about the observable signatures at HST resolution of a supermassive black hole in M32.
The black hole is expected to be associated with three kinematical features.
(1) The stellar rotation velocity should remain flat, or rise slightly, into the central resolution element.
(2) The stellar velocity dispersion should exhibit a sudden upturn at a distance of $\sim 0.2''-0.5''$ from the center, 
reaching a central value of $\sim 120$ \kms\ or greater depending on the mass of the black hole.
(3) The distribution of line-of-sight stellar velocities in the resolution element centered on the black hole should be strongly non-Gaussian, with extended, high-velocity wings.
Part of the predicted rise in the velocity dispersion near the center would be due to these wings; 
part to blending of the rotation curve from the two sides of the galaxy; 
and part to an intrinsic rise in the random velocities.

We see all three signatures of the black hole in the STIS data.
The velocity dispersion spike is most impressive; 
in terms of the usual parametrization $\sigma_0$
(the dispersion of the Gaussian core of the LOSVD),
the central measured value is $\sim 130$ \kms, 
while the value of $\sigma$ corrected for the non-Gaussian wings of the LOSVD is considerably greater, at least $175$ \kms, or $\sim 3$ times the value in the main body of M32.

A detailed description of the observations and the data reduction is given in \S 2.
In \S3 we describe the methods used to extract the stellar LOSVDs from the STIS spectra.
We carried out independent analyses based on two spectral deconvolution routines: 
the FCQ algorithm of R. Bender (1990), which is a Fourier method;
and the MPL routine of D. Merritt (1997), which is based on 
nonparametric function-estimation techniques.
The two algorithms gave consistent results for the low-order velocity moments of M32 (the mean and dispersion) but there were systematic differences in the recovered values of $h_4$, the parameter that measures symmetric deviations of the LOSVD from a Gaussian.
We argue that the $h_4$ values are consistent after taking into account the biases in the two methods; 
if our analysis is correct, the true values of $h_4$ in M32 are significantly greater than zero both within the region where the black hole's force is dominant and at larger radii.

A preliminary interpretation of the M32 kinematical data is presented in \S 4, where we address two basic questions:
Have we resolved the central spike in the velocity dispersions?
and: Is the shape of the central LOSVD consistent with what is expected for a nucleus containing a black hole?
We answer ``yes'' to both questions, 
although the resolution of the central spike may be marginal if the black hole mass is near the lower limit of the allowed range.
In any case, 
M32 is the first galaxy for which the imprint of the black hole's gravitation on the stellar velocities has been observed with a resolution comparable to that of the gas-dynamical studies (e.g. \cite{bow98}; \cite{fef99}). 

We show that both the velocity dispersion spike and the wings of the central LOSVD are consistent with the predictions of simple dynamical models containing black holes, with masses $M_h$ in the range $2\times 10^6\Msolar\lap M_h\lap 5\times 10^6\Msolar$.
More accurate estimates of $M_h$ will be presented in Paper II where the full kinematical data set will be compared to three-integral axisymmetric models.

We adopt a distance to M32 of $0.7$ Mpc; thus $1''$ corresponds to 3.25 pc. 

\clearpage
\section {Observations and Data Reduction}

M32 was observed on 1998 September 04 with STIS in the long-slit mode with
wavelength centered on the Ca II triplet feature near 8561\AA .
STIS is described by Woodgate et al. (1998) and its on-orbit performance by
Kimble et al. (1998).  The present data are part of a survey
of the nuclei of nearby galaxies 
being conducted by the STIS IDT (HST Program ID: 7566).  
The goal of the survey is to place stellar
kinematical constraints on the masses of nuclear black holes.
Two orbits of data with a total integration
time of 4,898 seconds were obtained.  The HST spacecraft tracking was operated
in fine lock with a reported jitter of no more than 3 mas rms or 11 mas peak-to-peak.  
The aperture was $52'' \times 0.1''$ with a position angle of 163$^{\circ}$,
coincident with the M32 isophotal major axis.  
The CCD data were read out in the unbinned mode.  
Spatial sampling at the focal plane was at every $0.05071''$,
corresponding to a 2-pixel optical resolution of about $0.115''$ FWHM.
Outside of $\pm 0.7''$, the data were binned spatially to
enhance the signal.  
The spectral resolution was approximately 38 km s$^{-1}$.  
M32 spectra were
obtained in the CR-split mode to assist with cosmic ray (CR) identification
and rejection.  
The location of the galaxy center in the image was moved by
approximately 4.5 rows along the aperture between the two orbits to ensure that
residual detector sensitivity variations, that may not be completely removed
from the data during reduction, are not mistaken for weak features.  
This form of dithering also assists with the identification of hot pixels in the CCD which do not rectify well.

A spectrum of the star HR 7615 was obtained with the same STIS aperture earlier in the program (Figure \ref{fig_spectra}).
This bright, K0 III giant was used as a template when inferring 
the LOSVD of the stars in M32 from the STIS spectra.  
Template spectra of a handful of other bright, cool stars of various spectral types were also obtained and used to test the sensitivity of the spectral deconvolution to template mismatch.  
A set of spectral images of HR 7615 were taken centered at $-0.05'', 0.00''$ and $+0.05''$ with respect to the centerline of the aperture
(i.e. offset along the dispersion direction).  
These data were added together using appropriate
weights to match the aperture illumination profile of M32.  
Each of these
spectral images has a slightly different shift in its velocity centroid
and the combined image provides a more accurate template for determining
the kinematics of an extended object such as a galaxy.  
Internal wavelength calibration images (``Wavecals'') 
as well as an internal continuum lamp image (``Flat field'') 
were taken for calibration purposes.  An image that
has had all of the instrumental response removed is said be a rectified image.
The contemporaneous flat field spectrum was obtained in the portion of the
orbit where HR 7615 was behind the earth.
The flat spectrum was taken through the $0.2''\times 0.09''$ aperture rather than through
the long aperture since the former is superior for removing the pixel-to-pixel
detector response of a stellar point source.  A contemporaneous flat field
was used to remove the internal fringing which is significant for
wavelengths greater than 7500 \AA\ and which changes over time (\cite{gbw97}).
We note that the fringing is far more serious for low dispersion spectra
especially for high S/N than it is for our data taken in medium dispersion.

Cosmic rays (CRs) account for approximately 20\% of the total signal and
contaminate approximately 5\% of the pixels in a typical exposure.  
CRs were identified and removed using the following procedure.  The centroids
in the cross-dispersion direction were determined for each rectified image
and the images shifted so that the galaxy core appeared on the same row.
CRs were identified and removed by comparing the flux in a given pixel to
the flux in the corresponding pixel in subsequent images.  
For each pixel, 
outlying values were rejected and excluded when the frames were averaged
together.  
Our data set for M32 included 4 raw images.  
Most pixels were found to have 4 frames contributing to their average values, 
while fewer pixels had 3 or less frames.  
Only 2 pixels within the central $2''$ had contributions from no frames.
Those pixels were assigned values representing the average of their adjacent
pixels.

The M32 data were reduced using two separate approaches:
1) ``Shift and Add,'' and 2) ``Frame by Frame.''  The latter relies
heavily on a standard software package called CALSTIS at Goddard Space
Flight Center.  The ``Shift and Add'' method
starts by removing the detector response using contemporaneous flat, bias,
and dark calibration files.  
CR hits are then removed using the procedure described previously.  
The frames are averaged together and the resulting 
frame is remapped to place the spectra from a single location along
the aperture onto a single row.  
As with most spectrographs, STIS produces spectra with S-shaped and 
pincushion distortions as well as spectra that are not aligned exactly 
with a row.  
A cubic interpolation was used to remap the spectra for later analysis.  
The remapping is not perfect,
with centroid errors of approximately 0.1 pixel rms.  
This level of accuracy was deemed adequate, 
although we are working to improve it.
In addition, the remapping produces very minor ($\lap 1\%$)
residual Moir\'e ripples in the data, which can be minimized but not completely
eliminated.  
Further work is also underway to measure and correct
for this residual affect.  
The strength of the ``Shift and Add'' technique is that it
preserves photometric accuracy.  
However it has the disadvantage of
introducing a subpixel image smearing since each frame is registered
to the nearest integer pixel.

The philosophy of the ``Frame by Frame'' method is to apply all calibration
corrections (including the remapping described above) to a single frame
before the resulting frames are added together.  
This approach has the
advantage of preserving the highest spatial resolution.  
However, a
substantial fraction of the pixels in each frame in this case are interpolated
values, potentially sacrificing some photometric precision.  
As noted above,
CRs contaminate approximately 5\% of the pixels in a 20 minute exposure.
After remapping, as many as 20\% of the pixels must be assigned a reduced
weight for the final averaging since a single pixel often gets remapped
partially into several pixels in the new image.\\

Fortunately, both approaches to the data reduction were found
to give very similar results.  
While the "Frame by Frame" method produced a higher apparent S/N ratio 
than the "Shift and Add" method, 
the former method is more prone to the introduction of systematic error.  
We adopted the more conservative approach of accepting
a somewhat higher variance rather than risk the introduction of a bias.
We therefore adopted the "Shift and Add" spectra for the analysis of the 
LOSVDs in the present study.  
Figure 1 shows spectra at several positions along the aperture.  
As noted above, we are continuing in our efforts to refine the data reduction 
techniques still further.

\clearpage
\section{Recovery of the Stellar Velocity Distribution}
\subsection{Method}

An observed spectrum $I(\lambda)$ is the convolution of the line-of-sight velocity distribution $N(V)$ of the stars within the aperture with the spectrum of a single star $T(\lambda)$:
\begin{equation}
I(\ln\lambda) = \int N(V)T(\ln\lambda-V/c)dV.
\end{equation}
The goal is to extract an estimate $\hat{N}(V)$ of the true stellar broadening function $N(V)$ given $I(\lambda)$ and $T(\lambda)$, both observed with the same instrument.
For $T(\lambda)$ we adopt the template spectrum of Figure \ref{fig_spectra}.

Two independent deconvolution routines were used.
The first algorithm, the ``Fourier Correlation Quotient'' (FCQ) method 
(\cite{ben90}; \cite{bsg94}),
constructs an estimate of the broadening function using Fourier techniques.
The FCQ routine differs from earlier Fourier algorithms (e.g. \cite{sar77}) in that the deconvolution is based on the template-galaxy correlation function rather than on the spectra themselves.
This approach is less sensitive to template mismatch (\cite{ben90}).
The second algorithm, ``Maximum Penalized Likelihood'' (MPL), 
finds $\hat{N}(V)$ as the solution to a penalized likelihood problem.
The MPL estimate of $N(V)$ is computed on a grid in $V$ in such a way as to optimize the fit of the convolved template to the galaxy spectrum, 
subject to a ``penalty'' that measures the lack of smoothness of $\hat{N}(V)$
(\cite{mer97}).

Both algorithms are nonparametric in the sense that no explicit constraints are placed on the functional form of $\hat{N}(V)$.
However they differ in two ways that are important for the current study.
The FCQ algorithm requires that the absorption lines in the template spectrum 
be narrow compared to the broadened lines of the galaxy spectrum, 
i.e. that the galaxy velocity dispersion be large compared to the instrumental resolution.
The MPL routine works well even when the galaxy velocity dispersion is small, as long as both template and galaxy spectra are observed at the same spectral resolution, 
at least in the case that the template star and galaxy have the same intrinsic absorption line properties (an assumption that will not be tested here).

The two algorithms differ also in the way they deal with the amplification of noise that accompanies the deconvolution.
The FCQ routine uses a Wiener filter to suppress high-frequency components of the template-galaxy correlation function $\tilde K_{T,G}$.
The degree of smoothing is determined by a factor, called here $W$,
which fixes the width of the Gaussian function used to model the low-frequency, or signal, component of $\tilde K_{T,G}$.
The choice $W=1$ corresponds to ``optimal'' filtering and larger values produce less smoothing; the FCQ
algorithm adopts a default value of $W=1$ but automatically increases $W$ (to a maximum of 1.3) if the recovered LOSVD shows evidence of significantly non-Gaussian wings.
In the MPL algorithm, the level of smoothing is determined by a factor $\alpha$ that multiplies the smoothness penalty function.
This penalty function defines any $N(V)$ that is Gaussian as ``smooth,'' 
regardless of its mean or dispersion, via Silverman's (1982) prescription; 
in the limit $\alpha\rightarrow\infty$, 
the MPL estimate of $N(V)$ is the Gaussian function which is most consistent, in a maximum-likelihood sense, with the galaxy spectrum.
There is no a priori way of computing the optimum value of $\alpha$ in the MPL algorithm, a point that we return to below.

The different effects of smoothing on the form of $\hat{N}(V)$ are illustrated in Figure \ref{fig_compare}, which shows estimates of the LOSVD in the central resolution element of M32 as computed by the two routines.
Both algorithms produce rapidly-fluctuating solutions when undersmoothed,
a consequence of the amplification of noise that always accompanies deconvolution.
The only significant difference in this regime is the non-negativity of the MPL estimates, 
a consequence of the logarithmic form of the penalty function (\cite{sil82}).
As the smoothing is increased, systematic differences begin to appear which are related to the different smoothing algorithms in the two codes.
Solutions obtained via MPL tend to be more robust with respect to the degree 
of smoothing, producing in the limit of large $\alpha$ a Gaussian fit.
However Figure \ref{fig_compare} suggests that estimates of certain quantities, e.g. the wings of the LOSVD, might depend sensitively on the choice of smoothing level in either algorithm.

Once an estimate of $N(V)$ has been obtained, 
various quantites related to the line-of-sight velocity distribution can be derived.
The simplest of these are the mean and rms velocities, which we denote by $\overline{V}$ and $\sigma$ respectively.
As is well known, both quantities are difficult to estimate for numerically-recovered LOSVD's 
since they are sensitively dependent on the form of $\hat{N}(V)$ at large velocities where this function is most poorly determined.
A standard alternative is to describe $\hat{N}(V)$ by a Gram-Charlier or Gauss-Hermite (GH) series, the product of a normalizing Gaussian with a sum of Hermite polynomials $H_i$, both expressed in terms of $(V-V_0)/\sigma_0$
(\cite{tht90}).
The parameters $V_0$ and $\sigma_0$ take the place of $\overline{V}$ and $\sigma$; 
while their definitions are to an extent arbitrary,
these parameters are typically determined by requiring the coefficients of $H_1$ and $H_2$, called $h_1$ and $h_2$, 
to be zero (\cite{ger93}; \cite{vdf93}).
Because $V_0$ and $\sigma_0$ describe the Gaussian core of the LOSVD, they are relatively insensitive to deviations of $\hat{N}(V)$ from Gaussianity at high velocities.
Information about these deviations is contained in the higher-order coefficients $h_3$, $h_4$ etc.;
$h_3$ measures asymmetries in $N(V)$ and $h_4$ measures the strength of symmetric, non-Gaussian wings.

The FCQ and MPL algorithms derive the GH parameters from $\hat{N}(V)$ in slightly different ways; details are given in Appendix A.

When applied to the STIS M32 spectra,
the two algorithms were found to give consistent results for the lowest moments of $N(V)$, i.e. $V_0$, $\sigma_0$ and $h_3$.
However the estimates of $h_4$ differed significantly at positions outside of the central $\sim 0.1''$.
The FCQ algorithm gave $-0.15\lap\hat{h}_4\lap 0$ at almost all positions; 
negative values of $h_4$ imply an $N(V)$ that falls off more sharply than a Gaussian at large velocities.
The MPL algorithm gave values for $h_4$  
in the range $0\lap h_4\lap 0.1$, almost all positive,
corresponding to LOSVD's with super-Gaussian wings.
Positive values of $h_4$ are expected near a black hole 
(\cite{baw76}; \cite{vdm94}) and are also characteristic of models with 
radially-anisotropic velocity distributions.

We discuss the origin of this discrepancy in Appendix B.
We believe that the primary reason for the systematic difference in $\hat{h}_4$ values is the low velocity dispersion of M32.
When a galaxy's velocity dispersion is comparable to the 
dispersion of the template star spectrum
($\sim 50$ \kms\ in the case of HR7615), 
the FCQ algorithm has difficulty recovering the true LOSVD
(Figures B2,3; \cite{bpn91}).
The $\hat{N}(V)$'s recovered by FCQ in this regime are more sharply truncated than the true $N(V)$'s, leading to systematically low estimates of $\hat{h}_4$.
For values of $\sigma_0$ and S/N comparable to those of M32 at $\sim 1''$, 
Figure B2 shows that the estimates of $h_4$ generated by FCQ depend only weakly on the true $h_4$, with a bias that approaches $-0.1$ for a true $h_4$ of 0.1.
The MPL algorithm suffers much less from this bias (Figures B3-5).

Bias in nonparametric function estimates can always be reduced 
by reducing the degree of smoothing (e.g. \cite{sil86}), 
which in the case of the FCQ algorithm means increasing $W$.
Figure B5c suggests that increasing $W$ from its default value of $1$ to 
values of $\sim 2$ can reduce the bias in FCQ estimates of $h_4$ by factors of $2$ or greater, even when $\sigma_0$ is as large as $100$ \kms .
We carried out this experiment with the STIS data; 
the results are shown in Figure \ref{fig_h4}.
The average $\hat{h}_4$ values recovered by FCQ in M32 are indeed dependent on $W$; a change in $W$ from $1$ to $1.5$ has the effect of increasing $\langle\hat{h}_4\rangle$ from $\sim -0.08$ to $\sim +0.08$.
The latter value is essentially identical to the mean value of $\hat{h}_4$ recovered via MPL.

We note that the dependence of $\hat{h}_4$ on $W$ could be due either to the suggested explanation, i.e. the need to include more frequency channels for dispersions close to the instrumental resolution, 
or alternatively to a mismatch between galaxy and template spectral properties in the wings of the lines which results in incorrect continuum subtraction.
We will not explore the second possibility here but note again that the Monte Carlo experiments in Appendix B suggest that values of $W$ of order 2 are appropriate even when there is no template mismatch.

The values of $\hat{h}_4$ recovered by MPL are also dependent on the value of the smoothing parameter $\alpha$ but much less so (cf. Figure \ref{fig_compare}), until $\alpha$ is made so large that the LOSVD is forced into a Gaussian shape. 
The Monte Carlo tests summarized in Figure B5b suggest that the bias in $\hat{h}_4$ as recovered by MPL is likely to be only of order $\sim -0.02$,
several times smaller than with FCQ.

We conclude that the values of $\hat{h}_4$ recovered by the two algorithms are consistent once their different biases are taken into account and that the values returned by MPL are likely to be more accurate.
Henceforth we adopt the MPL estimates.

The full set of GH parameters derived from the STIS spectra and their $1\sigma$ confidence intervals are given in Table 2 (FCQ) and Table 3 (MPL).
For radii $\lap 0.7''$ from the center the data were sampled at full resolution($\sim 0.05''$) while at larger radii they were binned spatially.
The sampling at small radii is fine enough that the data points are somewhat correlated; this was done to ensure that no information concerning the steep radial gradients of the profiles was lost.

\subsection{Results for M32} 

Figure \ref{fig_broad} presents LOSVDs computed via the MPL algorithm at positions separated by about $0.1''$ along the slit.
One expects these broadening functions to obey $N(V;R) = N(-V;-R)$, since for a point-symmetric galaxy, the velocity distributions should reverse after passing from one side of the galaxy to the other.
The LOSVDs of Fig. \ref{fig_broad} show approximately the expected symmetry.
The right-hand column of Figure \ref{fig_broad} plots mean broadening functions averaged over the two sides of the galaxy, $\overline{N}(V)={1\over 2}[N(V,R)+N(-V,-R)]$;
the central LOSVD has been symmetrized about $V=0$.

These broadening functions show clear and consistent deviations from Gaussian form, in two respects.
First, the central LOSVD exhibits strong super-Gaussian ``wings'' at high velocities.
These wings are possibly present also in some of the off-center LOSVDs although with lower amplitude.
Second, the off-center LOSVDs are asymmetric, with tails extending toward velocities opposite in sign to the mean velocity at each radius. 
These asymmetric tails are similar to those exhibited by a rotating system superimposed on a slowly-rotating bulge.

Systematic problems in the spectral deconvolution, e.g. template mismatch or incorrect continuum subtraction, can easily produce features like the wings and tails seen in the broadening functions of Figure \ref{fig_broad}.
Such errors in most cases would be expected to produce features located at the {\it same} velocity on both sides of the galaxy (e.g. \cite{bsg94})
and are therefore an unlikely explanation for the asymmetric tails seen in the off-center LOSVDs.
The strong wings seen in the central LOSVD might more plausibly be attributed to systematic errors.
However we found that the wings in the central LOSVD were robust; 
they appeared in both the MPL and FCQ estimates of $N(V)$ (though less clearly in the latter -- see Figure \ref{fig_compare}) and were relatively unaffected by changes in the assumed continuum level or slope.
We carried out MPL deconvolutions where the fit to the galaxy spectrum was restricted to the region around only one, or two, of the three calcium-triplet lines; these LOSVDs also exhibited strong wings.
We also tried using one of the other available STIS stellar templates; again the wings were only slightly affected.
(The adopted template, Figure 1, 
produced the best overall fit to the galaxy spectrum.)
Finally, we show in \S4.2 that the wings are consistent with those predicted by stars in the gravitational field of a supermassive black hole.

Figure \ref{fig_rotate} shows the Gauss-Hermite parameter $V_0$, 
a measure of the stellar rotation, in the inner arc second of M32.
Also plotted are $h_3$, the lowest, odd GH moment of the LOSVD, 
and the ``corrected'' rotation velocity, $V_{0,c}=V_0 + \sqrt{3}\sigma_0h_3$.
$V_{0,c}$ is a closer approximation than $V_0$ to the true mean line-of-sight velocity $\overline{V}$ (\cite{vdf93}).
The corrected rotation velocity is lower in absolute magnitude than $|V_0|$ due to the asymmetric wings of the LOSVD noted above.

The STIS rotation curve is consistent with earlier ground-based measurements (Figure \ref{fig_gb}) at radii $\gap 1''$ but with a larger peak value,
$\sim 60$ \kms.
Furthermore the rotation curve remains flat or slightly rising into smaller radii than seen heretofore, before falling at $R\lap 0.25''$ due to the blending of light from the two sides of the galaxy.
There is a suggestion of an east-west asymmetry in the rotation curve though the effect is probably not significant.

The $h_3$ profile is approximately antisymmetric about the center of M32, as expected in a relaxed galaxy.
$|h_3|$ reaches a maximum value of $\sim 0.05$ at $|R|\approx 0.3''$ and appears to gradually decline at larger radii.
This behavior is similar to that predicted in axisymmetric models (e.g. Figure 8 of \cite{deh95}) where $h_3$ remains essentially constant at radii outside the seeing disk.

The Gauss-Hermite parameter $\sigma_0$ is shown in Figure \ref{fig_dispersion}.
Also plotted is $h_4$, the lowest, even moment of the LOSVD, and the ``corrected'' velocity dispersion, $\sigma_{0,c}=\sigma_0(1+\sqrt{6}h_4)$; 
$\sigma_{0,c}$ is a closer approximation than $\sigma_0$ to the true rms velocity $\sigma$.
The velocity dispersion rises suddenly inside of $\sim 0.3''$, approximately the same radius at which the rotation curve begins to fall.
This coincidence suggests that at least part of the rise in $\sigma_0$ is due to averaging of the rotation velocity over the two sides of the galaxy near the center, which has the effect of converting a rotation into an apparent dispersion 
(\cite{ton87}).

The corrected velocity dispersion $\sigma_{0,c}$ rises well above $\sigma_0$ near the center due to the strong non-Gaussian wings of the LOSVD.
The central value of $\sigma_{0,c}$ is $\sim 175$ km s$^{-1}$; 
this should probably be interpreted as a lower limit since $h_4$ is only sensitive to the inner parts of the wings.
(We argue below, based on model fits, that the rms velocity in the central resolution element may be as high as $\sim 200 $ \kms.)
The ground-based data (Figure \ref{fig_gb}) are consistent with the STIS dispersions at radii $\gap 1''$ but fail to resolve the continued rise in $\sigma_0$ inside of $\sim 0.5''$.

Dynamical models (e.g. \cite{deh95}; \cite{qia95}) predict $h_4(R)$ profiles similar to that in Figure \ref{fig_dispersion} when observed with HST resolution:
a central maximum; a rapid drop, to small or negative values, at $R\approx 0.1''$; 
and a nearly constant value at larger radii.
The predicted drop at $\sim 0.1''$ is due to blending of the light from the two sides of the galaxy,
which broadens the low-velocity part of the LOSVD and lowers the observed $h_4$.
The predicted central value of $h_4$ depends strongly on the black hole mass and on the PSF;
our value, $\hat{h}_4\approx 0.14$, is larger than in the two studies just cited, but these studies were based on rather low assumed black hole masses, $M_h=1-2\times 10^6\Msolar$.
The true black hole mass is probably greater
(\cite{vdm98}).

The behavior of $\hat{h}_4$ at larger radii is surprising.
Previous observational studies (e.g. \cite{vdm94a}; \cite{bkd96}) have returned smaller estimates for $h_4$ in M32.
However we believe that these earlier results are not inconsistent with ours giving the difficulties involved with estimating this parameter.
The van der Marel (1994a) study was based on WHT observations with a much lower spatial resolution than the STIS data.
At radii $\lap 1''$, the value of $\hat{h}_4$ recovered by those authors was strongly affected by the PSF blending of the rotation curve discussed above,
yielding negative values in the central aperture. 
Outside of $\sim 2''$, van der Marel et al. found $\hat{h}_4$ to increase sharply to $\sim 0.03$ on both sides of the galaxy (their Figure 12).
Van der Marel (private communication) notes that the values of $\hat{h}_4$ derived from the WHT data depended sensitively on the choice of template spectrum and on the algorithm for continuum subtraction.
Using a single, best-fit template, $\hat{h}_4$ was found to lie between $\sim0.3$ and $\sim0.5$ throughout the inner $2''$;
the lower values of $\hat{h}_4$ in the published paper were derived using a spectral deconvolution routine that constructs an ``optimal'' template by linear superposition of a set of stellar spectra. 
In another ground-based study, Bender, Kormendy \& Dehnen (1996) applied the FCQ algorithm to CFHT data of higher spatial resolution and found $\hat{h}_4\approx 0.05$ inside of $0.2''$, gradually falling to $\sim 0$ at $\sim 1.0''$.
However the spectral resolution in this study was only $80$ \kms and the derivation of $h_3$ and $h_4$ correspondingly difficult;
as noted above, we also found $\hat{h}_4\approx 0$ from the STIS data using the FCQ algorithm and argued that these values were significantly negatively biased.

Although we believe that all of these studies are consistent with our conclusion that $h_4$ is significantly positive throughout the nucleus of M32,
we are less willing to strongly endorse the precise $\hat{h}_4$ values shown in Figure \ref{fig_dispersion}, due to the sensitive dependence of this parameter on the details of the spectral deconvolution algorithm, continuum subtraction, smoothing level, etc.
In Paper II we will construct dynamical models based on a range of assumed $h_4$ profiles in order to test the dependence of the inferred black hole mass on this parameter.

We may also compare our results to the van der Marel et al. (1997, 1998) HST/FOS measurements of $V_0$ and $\sigma_0$ (Figure \ref{fig_fos}).
The FOS measurements were taken through square apertures as small as $\sim 0.1''$ on a side, hence their spatial resolution is comparable to that of the STIS data.
However the FOS is a low spectral resolution instrument and not well suited to objects like M32 with a relatively low velocity dispersion; 
furthermore there are difficulties in positioning the FOS and these were probably the cause of the large point-to-point variations seen by van der Marel et al.
One advantage of STIS over FOS is the continuous spatial sampling which avoids potential errors in aperture placement.
We find a hint in the STIS data of the asymmetry seen in the FOS $\sigma_0(R)$ profile (a more rapid falloff on the west side).
The central FOS value of $\sigma_0$ seems significantly bigger than found here, and the FOS rotation velocities are systematically lower.

\clearpage
\section{Analysis}

The STIS data show what appear to be clear signatures of the gravitational influence of a massive compact object on the stellar velocity distribution within the central parsec of M32.
Here we use simple dynamical models to address the question of whether these observed features are consistent with the presence of a supermassive black hole.
Our aim is not to derive the best possible estimate of the black hole mass or the stellar velocity distribution -- those are the goals of Paper II -- but rather to address two, more basic issues concerning the interpretation of the data.

1. The velocity dispersion profile (Figs. \ref{fig_dispersion}, \ref{fig_gb}) exhibits a sudden upturn at a distance of $\sim 0.3''$ from the center, presumably due in part to the gravitational force from a massive compact object.
At roughly the same radius, the rotation curve falls (Figs. \ref{fig_rotate}, \ref{fig_gb}), presumably due to blending of light from opposite sides of the galaxy which are rotating in opposite directions.
The blending would also be expected to contribute to the rise in the observed velocity dispersion (\cite{ton87}), consistent with the fact that the upturn in $\sigma$ and the drop in $\overline{V}$ occur at roughly the same radius.
We would like to estimate the degree to which the velocity dispersion spike is a product of this blending, and the degree to which it is due to a real upturn in the stellar random velocities.
In other words: Do the STIS data resolve the black hole's sphere of influence?

2. The central LOSVD in M32 exhibits strong, super-Gaussian wings 
(Fig. \ref{fig_compare}, \ref{fig_broad}).
Such wings are a generic prediction of the black hole model 
(\cite{baw76}; \cite{vdm94}; \cite{deh95});
they result from stars on high-velocity orbits within the black hole's sphere of influence.
However systematic errors in the spectral deconvolution can also produce spurious features in the LOSVD's, particularly at large velocities where the form of the broadening function is only weakly constrained by the spectra.
We would like to verify that the inferred, non-Gaussian shape of the central LOSVD is consistent with that expected from the black hole model.

We note that M32 is the only galaxy so far to exhibit either a resolved central spike in the stellar velocity dispersions, or strong non-Gaussian wings in the nuclear LOSVD.
Either feature, if observed with sufficient spatial resolution and S/N, could independently place strong constraints on the mass of a central black hole.
We will in fact generate estimates of $M_h$ from our analyses of both features but we stress that the best estimates of $M_h$ can only come from more complete modelling based on the entire kinematical data set.

\subsection{The Velocity Dispersion Spike}
 
One expects to see a rise in the stellar velocities in a hot stellar system at a distance $\sim r_h$ from the black hole, where 
\begin{equation}
r_h = {GM_h\over \sigma_*^2},
\end{equation}
the black hole's ``radius of influence'' (\cite{pee72}).
Here $M_h$ is the black hole mass and $\sigma_*^2$ is the stellar velocity dispersion at $r>r_h$.
Setting $M_h = 10^6\Msolar$ and $\sigma_*=60$ km s$^{-1}$ (Figures \ref{fig_dispersion}, \ref{fig_gb}) gives $r_h \approx 1.2\ {\rm pc} \approx 0.3''$.
This value seems comfortably larger than the HST/STIS resolution but it is based on an assumed value of $M_h$ and furthermore it refers to the true 
radius whereas we observe the galaxy in projection, which tends to hide otherwise sharp features.
Modelling the spike therefore requires us to predict the true, projected velocity field of the galaxy in two dimensions on the plane of the sky, including both random and rotational velocities, and then to convolve it with the STIS PSF.

We begin by constructing solutions to the stellar hydrodynamical equations.
We assume the simplest possible axisymmetric model for the stars, in which the velocity dispersions are isotropic in the meridional plane $(\varpi,z)$, i.e. $\sigma_{\varpi}(\varpi,z)=\sigma_z(\varpi,z)\equiv\sigma(\varpi,z)$; the model is flattened by an inequality between $\sigma^2$ and the mean square azimuthal velocity $\overline{v_{\phi}^2}$.
The second moments of the internal velocity distribution are given by 
\begin{equation}
\nu\sigma^2 = \int_z^{\infty}\nu {\partial\Phi\over\partial z} dz, \ \ \ \ 
\nu\overline{v_{\phi}^2} = \nu\sigma^2+\nu\varpi{\partial\Phi\over\partial\varpi} + \varpi{\partial(\nu\sigma^2)\over\partial\varpi}
\end{equation}
where $\nu$ is the stellar number density, $v_{\phi}$ is the azimuthal velocity and $\Phi$ is the combined gravitational potential from the stars and the central black hole, $\Phi(\varpi,z)=\Phi_*(\varpi,z)-GM_h/r$
(\cite{mer99}).

We evaluated these expressions assuming a stellar density
\begin{equation}
\nu(\varpi,z) = \nu_0(m/b)^{\alpha}\left[1+(m/b)^2\right]^{\beta}\left[1+(m/c)^2\right]^{\gamma},\ \ \ \ m^2=\varpi^2+(z/q)^2,
\end{equation}
a parametrized form proposed by van der Marel et al. (1998); those authors found a good match between their model (observed edge-on) and M32 with the parameters $\alpha=-1.435, \beta=-0.423, \gamma=-1.298, b=0.55'', c=102.0'', q=0.73$ and $\nu_0=0.463\times 10^5 L_{\odot,V}$pc$^{-3}$.
The luminosity density $\nu_0$ is converted into a mass density $\rho_0$ by the factor $(M/L)(\Msolar/\Lsolar)$, with $M/L$ the mass-to-light ratio of the stars in solar units.

Given $\sigma(\varpi,z)$ and $\overline{v^2_{\phi}}(\varpi,z)$ as obtained from equations (3) and (4), 
the projected, line-of-sight, mean square velocity $\overline{V^2}$
is obtained by a density-weighted integration through the galaxy, assumed here to be edge-on:
\begin{equation}
\Sigma(R,Z)\overline{V^2}(R,Z) = 2\int_R^{\infty}\nu(\varpi,z)\left[\left(1-{R^2\over\varpi^2}\right)\sigma^2(\varpi,z) + {R^2\over\varpi^2}\overline{v_{\phi}^2}(\varpi,z)\right]{\varpi d\varpi\over\sqrt{\varpi^2-R^2}}
\end{equation}
(\cite{fil86}),
where $(R,Z)$ are coordinates on the plane of the sky and $R$ is measured parallel to the long axis of the galaxy's figure; $\Sigma(R,Z)$ is the stellar surface density.

Values of $\overline{V^2}$ were computed on a rectangular grid of $180\times 25$ locations with separations of $0.015''$ in $R$ and $0.02''$ in $z$.
These values were then convolved with the STIS PSF and averaged over the pixel area and the aperture after weighting by the model surface brightness.
The STIS PSF at $8500$\AA\  has a FWHM of $\sim 0.115''$.
The PSF is also slightly asymmetric (\cite{bow00}).
We ignored this slight asymmetry when carrying out the convolutions with our models.

The second velocity moments of models constructed in this way are uniquely determined by the two parameters $(M_h,M/L)$ that specify the potential.
Figure \ref{fig_contour} shows the goodness of fit of the models to the observed, mean square velocities; only data points within the inner $1.0''$ were used in evaluating $\chi^2$.
As estimates of $\overline{V^2}$, we took $V_{0,c}^2+\sigma_{0,c}^2$, where $V_{0,c}$ and $\sigma_{0,c}$ are the Gauss-Hermite parameters corrected by $h_3$ and $h_4$ respectively (Figures \ref{fig_rotate}, \ref{fig_dispersion}).
The best-fit model has $M_h\approx 3.2\times 10^6\Msolar$ and $M/L\approx 3.3$
with $\tilde{\chi^2}=0.64$; 
a $\tilde{\chi^2}$ of unity includes models with $M_h$ as small as $2.2 \times 10^6\Msolar$ and as large as $4.3 \times 10^6\Msolar$.

The degree of net rotation in our models may be adjusted by partitioning the azimuthal motions between streaming, $\overline{v_{\phi}}$, and dispersion, $\sigma_{\phi}$.
We followed the standard practice (\cite{sat80}) of making $\sigma^2_{\phi}$ a weighted average of $\sigma^2$ and $\overline{v_{\phi}^2}$, i.e.
\begin{equation}
\sigma_{\phi}^2 = k^2\sigma^2 + (1-k^2)\overline{v_{\phi}^2}.
\end{equation}
The parameter $k$ (assumed independent of position) 
may be varied between zero (corresponding to no streaming motions) and a maximum value, of order unity, at which $\sigma_{\phi}^2$ is forced below zero at some point in the meridional plane; $k=1$ yields an ``isotropic oblate rotator.''

When we add $k$ as a free parameter and require the models to fit the observed rotation and velocity dispersion profiles separately, the best-fit values of $(M_h,M/L)$ were nearly unchanged but $\tilde{\chi^2}$ increased to $3.7$ -- since the model is now being asked to fit twice as many data points with only one extra parameter.
The best-fit value of $k$ was found always to be close to $1.2$, 
implying slightly smaller $\sigma_{\phi}$ (i.e. greater rotation) 
than in an isotropic oblate rotator.
Figure \ref{fig_three} compares the data with the predicted profiles for 
$M_h=3.0\times 10^6\Msolar$, 
close to the best-fit value, 
and for $M_h=2.0$ and $4.0\times 10^6\Msolar$.

We draw the following conclusions from these comparisons.

1. The lowest-order moments of the line-of-sight velocity distribution in M32 are reasonably well fit near the center by our simple axisymmetric model, with a black hole mass $M_h\approx 3\times 10^6\Msolar$.
The rotation curve is best fit by a smaller mass ($\sim 1-2\times 10^6\Msolar$) and the velocity dispersions by a larger mass ($\sim 3-4\times 10^6\Msolar$);
if we require the models only to fit the mean square velocities, the fit is essentially perfect within the inner arc second.

2. The STIS observations probably come close to resolving the central upturn in the stellar velocity dispersions, 
which is predicted to occur at a projected radius of $\sim 0.1''$ for $M_h\sim 2\times 10^6\Msolar$ and $\sim 0.2''$ for $M_h\sim 4\times 10^6\Msolar$.

3. Smearing of the stellar rotation field probably accounts for only a small part of the observed upturn in the dispersions.

We note that Dehnen (1995) was able to improve the fit of his axisymmetric models to the M32 data then available by varying the ratio of rotational to non-ordered azimuthal motions, which in our models would correspond to varying $k$ with position.

If our simple model for the internal dynamics of M32 is approximately correct, Figures \ref{fig_contour} and \ref{fig_three} imply that the STIS data can place strong constraints on the mass of the central black hole.
However such a conclusion must await the results of the more complete modelling of Paper II.
That study may yield tighter constraints on $M_h$, due to the use of the full kinematical data set; or weaker constraints, due to the increased flexibility of general, anisotropic models.

\subsection{The Central Broadening Function}
 
The LOSVD near the projected center of a galaxy containing a black hole is expected to be very non-Gaussian due to high velocity stars orbiting near the central mass (\cite{baw76}).
If the stellar density follows a power law into the center, 
$\rho\propto r^{-\gamma}$, 
the LOSVD in an aperture containing the black hole has power-law wings, 
$N(V)\propto V^{2\gamma-7}$, $V\rightarrow\infty$ (\cite{vdm94}).
The amplitude of these wings depends on the ratio of black hole mass to stellar mass within the aperture, 
and on the slope $\gamma$ of the stellar density profile
(\cite{deh95}), among other factors.
The wings are most prominent in a galaxy, like M32, for which the stellar cusp is steep, $\gamma\approx 2$,
since a large fraction of the light near the projected center comes from the region near the black hole.

Here we ask whether the strong wings seen in the central M32 broadening function
(Figures \ref{fig_compare}, \ref{fig_broad}) 
are consistent with the black hole model.
To answer this question we must compute a stellar distribution function $f$
and integrate it over the two velocity components in the plane of the sky
(\cite{mer87}).
The result must then be smeared by the instrumental resolution.
We once again restrict ourselves to the simplest model which permits a meaningful comparison with the data; 
in this case, a spherical galaxy with an isotropic distribution function, $f=f(E)$.
Our model will ignore rotation, 
which acts to broaden the central LOSVD but leaves it symmetric. 
In any case we do not know what the contribution of rotation to the stellar velocity distribution is very near to the black hole.
For the stellar density profile we assume the spherical version of equation (4); 
$f(E)$ then follows from the standard formula (\cite{edd16})
and the projected velocity distribution is also straightforward to compute 
(\cite{mer93}).
We chose to fix $M/L$ for the stars at $3.5(\Msolar/\Lsolar)$ 
as the black hole mass was varied;  
this $M/L$ reproduces approximately the observed mean square velocities for $r\gg r_h$.

Figure \ref{fig_center} compares the central LOSVD in M32 to the broadening functions predicted by the spherical model, for black hole masses 
$M_h=(2.5,5.0,10.)\times 10^6\Msolar$.
The M32 LOSVD has been symmetrized about $V=0$.
The fit is quite reasonable for $M_h=5.0\times 10^6\Msolar$, 
although the high-velocity wings are better fit by still larger masses.
The model LOSVD for $M_h=5.0\times 10^6\Msolar$ has the parameters:
\begin{equation}
\sigma=178\ {\rm km/s}\ \ \ \ \sigma_0=133\ {\rm km/s}\ \ \ \ h_4=0.085.
\end{equation}
The first two numbers are essentially identical to the values inferred from the M32 data (Figure \ref{fig_dispersion});
$h_4$ is lower than, but consistent with, the M32 estimate ($0.14\pm 0.03$).

We conclude that the central LOSVD in M32 is consistent with that expected for a stellar nucleus containing a massive compact object, 
with a mass comparable to that found in the fit to the axisymmetric models.

\clearpage
\section{Summary}

We used HST and STIS to obtain stellar absorption line spectra near the center of M32 in a wavelength region centered on the Calcium triplet.
The spectra were analyzed using two independent spectral deconvolution routines; these gave fully consistent results except in the case of the Gauss-Hermite $h_4$ parameter, but we argued that the differences could be reconciled after taking into account the different biases of the two algorithms. 
The stellar rotation velocities in M32 are slightly higher than observed from the ground and remain constant into $\sim 0.25''$ from the center.
The velocity dispersions exhibit a clear spike beginning at approximately the same radius.
These two kinematical profiles are consistent with those predicted by simple axisymmetric models containing central black holes with masses in the range $2-5\times 10^6\Msolar$.
The stellar LOSVDs show significant deviations from Gaussian form as measured by the Gauss-Hermite parameters $h_3$ and $h_4$.
The central LOSVD is particularly non-Gaussian, exhibiting strong, high-velocity wings.
We showed that the amplitude of these wings is consistent with that predicted by simple models containing black holes with masses of order $3\times 10^6\Msolar$.

Detailed dynamical modelling of M32 based on these data and estimates of the black hole mass will be presented in Paper II.

\bigskip\bigskip

We thank W. Dehnen and R. van der Marel for helpful discussions.
This work was supported by NASA grants NAG 5-3158 and 
NAG 5-6037, by NSF grant AST 96-17088, and by STIS GTO funding.
Data presented here were based on observations with the NASA/ESA {\it Hubble Space Telescope}, obtained at the Space Telescope Science Institute, which is operated by the Association of Universities for Research in Astronomy, Inc. (AURA), under NASA contract NAS5-26555.

\clearpage

\centerline{\bf Appendix A}
\centerline{\bf Gauss-Hermite Moments}

The two spectral deconvolution algorithms described above yield nonparametric estimates $\hat{N}(V)$ of the stellar LOSVD.
Here we describe the methods used by the two algorithms to derive the GH moments from $\hat{N}(V)$. 

Let $N(X,Y;V)$ be the distribution of line-of-sight stellar velocities in the aperture centered at $(X,Y)$.
Define the GH moments of $N$ as
\beq
h_i(X,Y) = 2\sqrt{\pi}\int_{-\infty}^{\infty} N(X,Y;V)g(w)H_i(w)dV,
\eqnum{A1}
\eeq
where $H_i$ are the Hermite polynomials (as defined by \cite{ger93})
and the weight function
\beq
g(w) = {1\over\sqrt{2\pi}\gamma_0} e^{-w^2/2}, \ \ \ \ w=(V-V_0)/\sigma_0
\eqnum{A2}
\eeq
has three free parameters $(\gamma_0,V_0,\sigma_0)$.
Following van der Marel \& Franx (1993), we choose these parameters at every point $(X,Y)$ such that
\beq
h_0(X,Y)=1,\ \ \ \ h_1(X,Y)=h_2(X,Y)=0.
\eqnum{A3}
\eeq
These definitions impose the following implicit conditions on $(\gamma_0,V_0,\sigma_0)$:
\beq
\gamma_0 =\sqrt{2}\sigma_0\int_{-\infty}^{\infty} N(V)e^{-w^2/2}dw,
\eqnum{A4a}
\eeq
\beq
0 = \int_{-\infty}^{\infty} N(V)e^{-w^2/2}\ w\ dw,\eqnum{A4b}
\eeq
\beq
0 = \int_{-\infty}^{\infty} N(V)e^{-w^2/2}(2w^2-1)\ dw.\eqnum{A4c}
\eeq
The relations (A4) define a nonlinear minimization problem with solutions $(\gamma_0,V_0,\sigma_0)$ given $N(V)$.

The MPL algorithm (\cite{mer97}) derives the three parameters in just this way, using the NAG routine {\tt E04FDF} to minimize the sum
$(h_0-1)^2 + h_1^2 + h_2^2$ as a function of $(\gamma_0,V_0,\sigma_0)$.
The higher-order GH moments are then derived using equation (A1), by numerical integration over $\hat{N}(V)$.

Most spectral deconvolution algorithms of which we are aware
derive the parameters $(\gamma_0,V_0,\sigma_0)$ in a different way.
The LOSVD is compared to the trial function
\beq
{\cal N}(V) = {\gamma_0\over\sqrt{2\pi}\sigma_0}e^{-w^2/2}\left[1 + \sum_{j=3}^{j_{max}} h_jH_j(w)\right]
\eqnum{A5}
\eeq
where $j_{max}$ is the index of the highest GH moment fitted to $\hat{N}(V)$; typically $j_{max}=4$.
The integrated square deviation between $\hat{N}(V)$ and $\cal{ N(V)}$ is then minimized by varying the $(j_{max}+1)$ free parameters $(\gamma_0,V_0,\sigma_0,h_3,h_4,...,h_{j_{max}})$.
This is the technique used by the FCQ algorithm.

A theorem (\cite{myl08}) guarantees the equivalence of the two approaches if $j_{max}=\infty$ in equation (A5) (\cite{vdf93}).
However if $j_{max}\ne\infty$, and if the input $N(V)$ can not be precisely represented by a finite GH series with $j\le j_{max}$, the results given by the two algorithms will differ.
For instance, in attempting to represent an $N(V)$ having $h_6\ne 0$ using $j_{max}=4$, the FCQ algorithm will adjust $\sigma_0$ and $h_4$ to incorrect values in order to better fit the high-velocity wings of the profile with the limited number of terms allowed to it.
This is illustrated in Figure A1, which shows the values of $\sigma_0$ and $h_4$ generated by the second algorithm, $\hat\sigma_0$ and $\hat h_4$, compared to the true values for an input $N(V)$ with $h_4=0.15$ and nonzero $h_6$:
\beq
N(V) = {1\over\sqrt{2\pi}}e^{-V^2/2}\left\{1 + 0.15H_4(V) + h_6H_6(V)\right\}.
\eqnum{A6}
\eeq
For $|h_6|\gap 0.1$, the errors in $\sigma_0$ and $h_4$ as derived from the second algorithm are $\gap 15\%$ and $\gap 20\%$ respectively.

\clearpage
\centerline{\bf Appendix B}
\centerline{\bf Performance Evaluation of the FCQ and MPL Algorithms}

Here we compare the performance of the FCQ and MPL algorithms given simulated data.
Our primary goal is to understand the source of the systematic offset of $h_4$ values as derived from the M32 spectra by the two algorithms (\S3.1).
Two independent sets of tests were carried out, 
the first by R. Bender and the second by D. Merritt.
All tests were based on synthesized galaxy spectra generated 
from the STIS template spectrum (Figure \ref{fig_spectra}) 
by convolving it with an assumed $N(V)$ and adding noise.

The first set of tests addressed the accuracy of
FCQ estimates when the galaxy velocity dispersion is low.
It is well known that the accuracy of FCQ begins to fall off when the galaxy velocity dispersion becomes comparable to the dispersion of the template spectrum (e.g. \cite{bpn91}). 
Figure B1 shows values of $\hat{\sigma}_0$ recovered by FCQ given 
a Gaussian-broadened template spectrum and thirty random noise realizations.
The default value ($W\approx 1$) of the smoothing parameter was used.
There is a positive bias in the estimated values
beginning at $\sigma_0\approx 100$ \kms;
the bias increases with decreasing $\sigma_0$ becoming significant for $\sigma_0\approx 50$ \kms.
The bias is only weakly dependent on S/N.
This bias in the estimation of $\sigma_0$ is unlikely to be important for the nucleus of M32 where $\sigma_0\gap 100$\kms.

Figure B2 shows the performance of FCQ at recovering $h_4$.
The template spectrum was broadened using an $N(V)$ of the form
\beq
N_1(V) = {1\over\sqrt{2\pi}}e^{-V^2/2\sigma_0^2}\left\{1 + h_4H_4(V/\sigma_0)\right\}
\eqnum{B1}
\eeq
with various values of $\sigma_0$ and $h_4$.
Figure B2 reveals significant biases in $\hat{h}_4$ for $\sigma_0\lap 100$ \kms, even when S/N is as great as 100.
When $\sigma_0\approx 50$ \kms\ and S/N $\approx 30$, 
characteristic of M32 at $\sim 1''$, 
the bias in $h_4$ is $\sim -0.1$ for an input $h_4$ of $\sim 0.1$.

The second set of tests compared the performance of the
FCQ and MPL algorithms on galaxy spectra generated from the broadening function
\beq
N_2(V) = {1\over\pi\sigma}{1\over 1+(V/\sigma)^2},
\eqnum{B2}
\eeq
a Lorentzian function, with $\sigma=100$ \kms.
This LOSVD is qualitatively similar to what is expected in a black-hole cusp, 
with $N\sim V^{-2}$ high-velocity wings.
The non-trivial GH parameters are
\beq
\gamma_0=0.76986\ \ \ \ \sigma_0=108.07\ {\rm km\ s}^{-1} \ \ \ \ h_4=0.14546\ \ \ \ h_6=0.01850.
\eqnum{B3}
\eeq
Figure B3 shows mean estimates of $N(V)$ obtained using the two algorithms for 100 random realizations of the noise.
The smoothing parameter in both algorithms was adjusted to minimize the mean square error of $\hat{N}(V)$ (as defined below) 
for each value of S/N.
There is a greater bias in the FCQ estimates, 
as well as a persistent ``ringing'' at high velocities.

Figure B4 plots the mean integrated square error (MISE) and integrated square bias (ISB) of the recovered broadening functions as functions of S/N; 
in the case of the FCQ algorithm, 
the integrated errors are shown both for the optimal choice of smoothing parameter $W_{opt}$ that minimizes the MISE, 
as well as for the value chosen by the algorithm ($\sim 1.3$).
The MISE of an estimate $\hat{f}(x)$ is defined as
\beq
{\rm MISE}\left[\hat{f}(x)\right] = E\int\left\{\hat{f}(x) - f(x)\right\}^2dx 
\ \ \ \ \ \ \ \ \eqnum{B4a}
\eeq
\beq
\ \ \ \ \ \ \ \ \ \ = \int\left\{E\hat{f}(x) - f(x)\right\}^2dx + 
\int \left(E\left\{\hat{f}^2(x)\right\} - 
E\left\{\hat{f}(x)\right\}^2\right) dx \eqnum{B4b}
\eeq
\beq
= {\rm ISB}\left[\hat{f(x)}\right] + 
{\rm IV}\left[\hat{f(x)}\right], \ \ \ \ \ \ \eqnum{B4c}
\eeq
the sum of the integrated square bias ISB and the integrated variance IV 
(\cite{sil86}); here $E$ denotes the expectation value, i.e. the average over many random realizations of the noise.
The MISE and ISB displayed in Fig. B4 were divided by the normalizing factor 
$\int \left[N_2(V)\right]^2 dV$.

The MISE of the MPL estimates falls roughly as a power law, 
${\rm MISE}[\hat{N}(V)]\sim (S/N)^{-1.3}$, 
close to the asymptotic $(S/N)^{-1}$ of parametric estimators.
Approximately $1/2$ of the total square error comes from the bias and $1/2$ from the variance.
In the case of the FCQ algorithm, the MISE behaves in a more complicated way with S/N, at first falling with S/N then appearing to level off for S/N $\gap 50$.
This levelling off is a consequence of the low-velocity-dispersion bias of FCQ discussed above.
For the FCQ estimates, the bulk of the MISE is due to the variance;
adjusting the smoothing parameter primarily affects the bias and has little effect on the MISE.
For S/N $\approx 20$, the mean square error of the optimal FCQ estimate is a factor $\sim 3$ greater than that of the MPL estimates.

The bias in $\hat{N}(V)$ is in the direction of wider and more steeply truncated functions, particularly in the case of the FCQ estimates (Fig. B3).
This bias in $\hat{N}(V)$ is consistent with the negative bias found above 
in estimates of $h_4$.
Figure B5 compares the ability of the two algorithms to recover $h_4$ from the Lorentzian $N_2(V)$.
Plotted there are the mean square error (MSE) and bias in estimates of $h_4$ from 100 random noise realizations;
the MSE is defined, for any estimated parameter $\hat{P}$, as
\beq
{\rm MSE}(\hat{P}) = E\left\{\hat{P} - P\right\}^2,
\eqnum{B5}
\eeq
which can also be decomposed into contributions from the squared bias SB and the variance V:
\beq
\ \ \ \ {\rm MSE}(\hat{P}) = \left(E\hat{P} - P\right)^2 + \left(E\left\{\hat{P}^2\right\} - E\left\{\hat{P}\right\}^2\right) \eqnum{B6a}
\eeq
\beq
= {\rm SB}(\hat{P}) + {\rm V}(\hat{P}). \ \ \ \ \ \ \ \ \eqnum{B6b}
\eeq
The MSE of estimates obtained with the MPL algorithm again varies roughly as a power law, ${\rm MSE}(\hat{h_4})\sim (S/N)^{-1.5}$.
The bias in the MPL estimates is always negative, i.e. in the direction of more Gaussian $N(V)$'s;
for $S/N\sim 20$, this bias is a modest $\sim -0.03$,
dropping to $\lap -0.01$ for S/N $= 100$.

The FCQ estimates of $h_4$ show a considerably greater error, 
both in the bias and the variance.
Two sets of FCQ estimates were made: 
first using the default value of the smoothing parameter returned by the code, $W\approx 1.3$; 
and second using the optimum value $W_{opt}$ that minimized the MSE of the $h_4$ estimates at each S/N.
For the default value of $W$, 
the algorithm returns mean estimates of $h_4$ that lie in the range $0.03-0.05$ for all values of S/N $\ge 10$, an average error of $\sim 70\%$.
However the optimum smoothing parameter for the recovery of $h_4$ was found to vary strongly with S/N, from $W_{opt}\sim 0.5$ for S/N $=5$ to  $W_{opt}\sim 2$ for S/N $=100$ (Fig. B5c).
Nevertheless a substantial bias remains when $W_{opt}$ is used, 
of order $\sim -0.05$ even for S/N $=50-100$.
These biases are larger than found above using a more Gaussian $N(V)$ with smaller $h_4$ and suggest that FCQ estimates of $h_4$ may be substantially biased even for $\sigma_0$ as large as $\sim 100$ \kms\ when the true $N(V)$ is sufficiently non-Gaussian.

\clearpage


\begin{deluxetable}{ll}
\tablenum{1}
\tablecaption{Observational Setup
              \label{t:Setup}}
\tablehead{ }
\startdata
Gain (e$^-$/ADU) & 1.0 \\
Wavelength coverage & 8275\AA\ -- 8847\AA\ \\
Reciprocal dispersion (\AA\ pixel$^{-1}$) & 0.56 \\
Aperture & $52''\times 0.1''$ \\
Comparison line FWHM (pixels) & 2.0 \\
$R=\lambda/\Delta\lambda$ & 7644 \\
Instrumental dispersion ($\sigma_I$)(\kms) & 17.1 \\
Spatial scale ($''$ pixel$^{-1}$) & 0.05071 \\

\enddata
\end{deluxetable}

\clearpage


\begin{deluxetable}{ccccccccccccc}
\tablenum{2}
\tablecaption{M32 Kinematics as Derived via FCQ
              \label{t:kinFCQ}}
\tablehead{
\colhead{$R$} & \colhead{$V_0$} & \colhead{$\Delta V_0$} & 
\colhead{$\sigma_0$} & \colhead{$\Delta\sigma_0$} &
\colhead{$h_3$} & \colhead{$\Delta h_3$} & \colhead{$h_4$} & 
\colhead{$\Delta h_4$} & \colhead{$V_{0,c}$} & \colhead{$\Delta V_{0,c}$} & 
\colhead{$\sigma_{0,c}$} & \colhead{$\Delta\sigma_{0,c}$}
\\
\colhead{(1)} & \colhead{(2)} & & \colhead{(3)} & & \colhead{(4)} & &
\colhead{(5)} & & \colhead{(6)} & & \colhead{(7)} &
}
\startdata

 4.497 &  25.2 &  32.2 &  67.2 &  19.0 &   0.110 &  0.436 &  -0.289 &  0.436 &  38.0 &  60.2 &  19.6 &  75.5 \\
 4.000 &  49.7 &  31.5 &  97.5 &  32.2 &   0.053 &  0.294 &  -0.055 &  0.294 &  58.7 &  58.9 &  84.3 &  77.3 \\
 3.465 &  55.5 &  23.6 &  60.7 &  13.9 &  -0.063 &  0.353 &  -0.230 &  0.353 &  48.8 &  44.0 &  26.5 &  54.9 \\
 2.983 &  54.4 &  20.6 &  82.2 &  18.9 &   0.017 &  0.227 &  -0.090 &  0.227 &  56.8 &  38.3 &  64.1 &  49.6 \\
 2.546 &  35.0 &  18.8 &  60.1 &  14.0 &   0.040 &  0.284 &  -0.148 &  0.284 &  39.1 &  35.0 &  38.3 &  44.4 \\
 2.173 &  24.1 &  25.1 & 103.3 &  18.7 &  -0.052 &  0.221 &  -0.148 &  0.221 &  14.8 &  46.9 &  65.9 &  59.4 \\
 1.842 &  35.9 &  17.2 &  64.0 &  13.0 &  -0.058 &  0.245 &  -0.144 &  0.245 &  29.5 &  32.2 &  41.4 &  40.8 \\
 1.562 &  61.5 &  17.2 &  88.8 &  15.6 &  -0.074 &  0.176 &  -0.093 &  0.176 &  50.1 &  32.2 &  68.6 &  41.5 \\
 1.336 &  42.8 &  13.7 &  76.9 &  13.5 &  -0.013 &  0.162 &  -0.067 &  0.162 &  41.0 &  25.6 &  64.3 &  33.4 \\
 1.159 &  34.4 &  15.4 &  81.6 &  13.9 &   0.010 &  0.171 &  -0.095 &  0.171 &  35.8 &  28.6 &  62.6 &  37.0 \\
 1.034 &  53.6 &  15.3 &  82.9 &  12.6 &   0.059 &  0.168 &  -0.122 &  0.168 &  62.1 &  28.6 &  58.1 &  36.6 \\
 0.932 &  52.8 &  13.5 &  74.4 &  11.2 &  -0.024 &  0.165 &  -0.120 &  0.165 &  49.7 &  25.2 &  52.5 &  32.3 \\
 0.831 &  65.7 &  12.9 &  72.2 &  10.4 &  -0.078 &  0.163 &  -0.129 &  0.163 &  55.9 &  24.2 &  49.4 &  30.8 \\
 0.729 &  69.6 &  11.2 &  77.1 &  11.3 &  -0.001 &  0.132 &  -0.057 &  0.132 &  69.4 &  20.9 &  66.3 &  27.4 \\
 0.654 &  62.4 &  12.6 &  72.5 &  11.0 &   0.018 &  0.158 &  -0.105 &  0.158 &  64.7 &  23.5 &  53.9 &  30.3 \\
 0.603 &  57.5 &  12.6 &  79.0 &   8.1 &  -0.055 &  0.144 &  -0.180 &  0.144 &  50.0 &  23.4 &  44.2 &  29.2 \\
 0.553 &  56.8 &  13.0 &  75.3 &  11.1 &  -0.014 &  0.157 &  -0.112 &  0.157 &  55.0 &  24.3 &  54.7 &  31.2 \\
 0.502 &  58.8 &  10.7 &  72.4 &  10.1 &   0.019 &  0.135 &  -0.083 &  0.135 &  61.1 &  20.1 &  57.7 &  26.1 \\
 0.451 &  58.8 &  11.6 &  81.1 &  11.2 &   0.000 &  0.130 &  -0.074 &  0.130 &  58.8 &  21.6 &  66.4 &  28.2 \\
 0.401 &  65.2 &  10.4 &  76.0 &  10.2 &  -0.089 &  0.124 &  -0.065 &  0.124 &  53.5 &  19.4 &  63.9 &  25.3 \\
 0.350 &  58.8 &  10.0 &  86.0 &  10.2 &  -0.067 &  0.106 &  -0.055 &  0.106 &  48.8 &  18.7 &  74.4 &  24.6 \\
 0.299 &  52.9 &   9.6 &  89.3 &  10.3 &  -0.072 &  0.098 &  -0.037 &  0.098 &  41.8 &  18.0 &  81.2 &  23.8 \\
 0.248 &  60.6 &   8.4 &  77.4 &   8.9 &  -0.041 &  0.099 &  -0.044 &  0.099 &  55.1 &  15.7 &  69.1 &  20.8 \\
\tablebreak
 0.198 &  53.4 &   8.1 &  90.4 &   8.3 &  -0.049 &  0.082 &  -0.057 &  0.082 &  45.7 &  15.2 &  77.8 &  20.0 \\
 0.147 &  40.8 &   9.2 &  99.4 &   8.9 &   0.038 &  0.084 &  -0.072 &  0.084 &  47.3 &  17.2 &  81.8 &  22.4 \\
 0.096 &  38.9 &  10.2 & 112.2 &  11.2 &   0.012 &  0.083 &  -0.030 &  0.083 &  41.2 &  19.1 & 104.0 &  25.4 \\
 0.046 &  36.3 &   8.7 & 123.0 &  10.8 &  -0.009 &  0.064 &   0.019 &  0.064 &  34.4 &  16.2 & 128.7 &  22.1 \\
-0.005 &  -2.9 &   9.0 & 136.6 &  12.6 &  -0.013 &  0.060 &   0.075 &  0.060 &  -6.0 &  16.8 & 161.7 &  23.8 \\
-0.056 & -24.6 &   9.6 & 124.2 &  10.1 &   0.032 &  0.071 &  -0.045 &  0.071 & -17.7 &  18.1 & 110.5 &  23.9 \\
-0.106 & -34.4 &   9.2 & 104.2 &   9.8 &   0.070 &  0.081 &  -0.041 &  0.081 & -21.7 &  17.3 &  93.8 &  22.9 \\
-0.157 & -40.9 &   9.9 &  98.3 &  11.5 &   0.087 &  0.092 &  -0.007 &  0.092 & -26.1 &  18.6 &  96.6 &  25.0 \\
-0.208 & -53.1 &   9.9 &  90.8 &  10.4 &   0.037 &  0.100 &  -0.045 &  0.100 & -47.3 &  18.6 &  80.8 &  24.6 \\
-0.259 & -64.0 &   9.9 &  87.3 &  10.6 &   0.013 &  0.103 &  -0.036 &  0.103 & -62.0 &  18.4 &  79.6 &  24.5 \\
-0.309 & -62.6 &   9.0 &  76.4 &   8.5 &   0.052 &  0.107 &  -0.079 &  0.107 & -55.7 &  16.8 &  61.6 &  21.8 \\
-0.360 & -63.4 &  10.2 &  80.8 &  10.8 &   0.045 &  0.114 &  -0.042 &  0.114 & -57.1 &  18.9 &  72.5 &  25.0 \\
-0.411 & -50.9 &  12.0 &  90.4 &  13.8 &   0.038 &  0.120 &  -0.010 &  0.120 & -44.9 &  22.3 &  88.2 &  30.0 \\
-0.461 & -57.6 &  11.6 &  70.5 &  11.0 &   0.016 &  0.150 &  -0.078 &  0.150 & -55.7 &  21.7 &  57.0 &  28.2 \\
-0.512 & -60.1 &  10.8 &  63.5 &   9.0 &   0.033 &  0.155 &  -0.118 &  0.155 & -56.5 &  20.2 &  45.1 &  25.9 \\
-0.563 & -55.0 &  13.5 &  83.4 &  12.1 &   0.092 &  0.147 &  -0.096 &  0.147 & -41.7 &  25.2 &  63.8 &  32.5 \\
-0.614 & -51.5 &  15.5 &  93.2 &  15.2 &   0.047 &  0.151 &  -0.067 &  0.151 & -43.9 &  28.9 &  77.9 &  37.8 \\
-0.664 & -51.4 &  13.4 &  66.4 &  11.6 &   0.053 &  0.183 &  -0.106 &  0.183 & -45.3 &  25.0 &  49.2 &  32.1 \\
-0.715 & -58.4 &  13.2 &  78.0 &  10.7 &   0.173 &  0.154 &  -0.126 &  0.154 & -35.0 &  24.8 &  53.9 &  31.5 \\
-0.766 & -46.2 &  18.0 &  87.3 &  17.3 &   0.189 &  0.188 &  -0.076 &  0.188 & -17.6 &  34.1 &  71.0 &  43.9 \\
-0.841 & -31.8 &  12.2 &  77.9 &  10.1 &  -0.079 &  0.142 &  -0.119 &  0.142 & -42.4 &  22.7 &  55.2 &  29.1 \\
-0.943 & -51.4 &  13.4 &  77.1 &  11.1 &  -0.009 &  0.158 &  -0.121 &  0.158 & -52.6 &  25.0 &  54.3 &  32.0 \\
\tablebreak
-1.044 & -44.9 &  13.5 &  74.9 &  13.8 &   0.079 &  0.164 &  -0.054 &  0.164 & -34.6 &  25.3 &  65.0 &  33.2 \\
-1.145 & -48.6 &  14.9 &  67.0 &  11.8 &   0.037 &  0.202 &  -0.132 &  0.202 & -44.3 &  27.8 &  45.3 &  35.4 \\
-1.247 & -54.2 &  15.0 &  73.8 &  14.1 &   0.064 &  0.184 &  -0.082 &  0.184 & -46.0 &  27.9 &  59.0 &  36.3 \\
-1.372 & -51.8 &  23.3 & 109.2 &  23.9 &  -0.044 &  0.194 &  -0.053 &  0.194 & -60.1 &  43.5 &  95.0 &  57.2 \\
-1.549 & -69.3 &  17.1 &  83.9 &  12.4 &   0.180 &  0.186 &  -0.154 &  0.186 & -43.1 &  32.2 &  52.3 &  40.5 \\
-1.778 & -52.1 &  16.9 &  75.6 &  15.4 &  -0.075 &  0.204 &  -0.092 &  0.204 & -61.9 &  31.7 &  58.5 &  40.9 \\
-2.052 & -56.6 &  16.3 &  81.1 &  13.5 &   0.077 &  0.183 &  -0.121 &  0.183 & -45.8 &  30.5 &  57.1 &  39.0 \\
-2.382 & -66.0 &  18.0 &  78.2 &  19.2 &   0.172 &  0.209 &  -0.039 &  0.209 & -42.8 &  34.0 &  70.7 &  44.4 \\
-2.765 & -54.6 &  29.4 & 102.9 &  29.4 &   0.048 &  0.259 &  -0.061 &  0.259 & -46.1 &  54.8 &  87.5 &  71.7 \\
-3.185 & -53.9 &  21.7 &  62.1 &  17.2 &   0.129 &  0.317 &  -0.131 &  0.317 & -40.1 &  40.6 &  42.2 &  51.5 \\
-3.673 & -70.0 &  51.5 & 167.0 &  55.9 &  -0.003 &  0.280 &  -0.033 &  0.280 & -70.8 &  96.0 & 153.5 & 127.6 \\
-4.206 & -46.9 &  25.2 &  30.0 &  33.3 &   0.039 &  0.763 &   0.047 &  0.763 & -44.9 &  47.0 &  33.5 &  65.3 \\
-4.722 & -67.4 &  32.1 &   6.5 &  53.7 &  -0.295 &  4.505 &   0.167 &  4.505 & -70.7 &  65.8 &   9.1 &  92.0 \\

\enddata
\tablecomments{
(1) Distance from center of M32 in arc seconds.
(2) Rotation parameter $V_0$ in \kms.
(3) Dispersion parameter $\sigma_0$ in \kms.
(4) LOSVD skewness parameter $h_3$.
(5) LOSVD kurtosis parameter $h_4$.
(6) $V_{0,c}=V_0 + \sqrt{3}\sigma_0 h_3$, an estimate of the true mean line-of-sight velocity.
(7) $\sigma_{0,c}=\sigma_0(1 + \sqrt{6}h_4)$, an estimate of the true line-of-sight velocity dispersion.
}
\end{deluxetable}

\clearpage


\begin{deluxetable}{ccccccccccccc}
\tablenum{3}
\tablecaption{M32 Kinematics as Derived via MPL
              \label{t:kinMPL}}
\tablehead{
\colhead{$R$} & \colhead{$V_0$} & \colhead{$\Delta V_0$} & 
\colhead{$\sigma_0$} & \colhead{$\Delta\sigma_0$} &
\colhead{$h_3$} & \colhead{$\Delta h_3$} & \colhead{$h_4$} & 
\colhead{$\Delta h_4$} & \colhead{$V_{0,c}$} & \colhead{$\Delta V_{0,c}$} & 
\colhead{$\sigma_{0,c}$} & \colhead{$\Delta\sigma_{0,c}$}
\\
\colhead{(1)} & \colhead{(2)} & & \colhead{(3)} & & \colhead{(4)} & &
\colhead{(5)} & & \colhead{(6)} & & \colhead{(7)} &
}
\startdata
 4.559 & 18.9 & 12.8 & 51.0 & 10.4 &  0.176 & 0.140 & 0.022 & 0.033 & 37.9 & 19.7 & 53.7 & 12.1\\
 4.184 & 70.2 & 18.9 & 97.2 & 30.0 & -0.235 & 0.155 & 0.040 & 0.079 & 7.15 & 51.4 & 106.7 & 31.4 \\
 3.545 & 47.3 &  8.4 & 47.9 &  6.8 &  0.054 & 0.102 & 0.008 & 0.032 & 56.9 & 10.1 & 48.8 & 8.9\\
 3.068 & 54.4 &  8.1 & 60.3 & 12.7 & -0.054 & 0.098 & 0.094 & 0.033 & 49.5 & 14.3 & 74.2 & 16.4\\
 2.632 & 44.8 &  6.7 & 45.4 &  7.0 & -0.008 & 0.111 & 0.028 & 0.018 & 35.4 & 14.1 & 48.6 & 9.6\\
 2.247 & 27.2 &  9.8 & 70.0 & 10.1 &  0.043 & 0.094 & 0.111 & 0.060 & 29.0 & 12.9 & 89.0 & 14.4\\
 1.922 & 37.6 &  7.8 & 62.7 &  9.0 & -0.102 & 0.080 & 0.048 & 0.031 & 26.9 & 11.4 & 70.0 & 10.1\\
 1.648 & 71.8 &  7.9 & 64.6 & 12.6 & -0.124 & 0.076 & 0.080 & 0.027 & 60.2 & 16.7 & 77.3 & 16.9\\
 1.415 & 46.4 &  6.9 & 54.8 &  7.1 & -0.008 & 0.079 & 0.069 & 0.023 & 46.1 & 10.0 & 64.1 & 9.5\\
 1.242 & 44.7 &  7.2 & 64.4 &  8.6 & -0.079 & 0.091 & 0.077 & 0.026 & 37.6 & 10.7 & 76.5 & 11.6\\
 1.111 & 49.1 &  8.1 & 87.5 & 11.2 &  0.070 & 0.079 & 0.037 & 0.047 & 62.7 & 13.2 & 95.5 & 11.8\\
 1.009 & 58.2 &  7.2 & 72.7 & 11.2 & -0.144 & 0.087 & 0.104 & 0.041 & 37.1 & 13.9 & 91.1 & 14.1\\
 0.908 & 71.7 &  4.6 & 49.6 &  6.5 & -0.034 & 0.072 & 0.036 & 0.012 & 69.3 & 6.7 & 54.0 & 8.5\\
 0.806 & 69.1 &  4.9 & 68.7 &  8.3 &  0.000 & 0.070 & 0.082 & 0.023 & 70.6 & 8.2 & 82.6 & 8.7\\
 0.705 & 49.5 &  7.0 & 67.4 &  8.7 &  0.138 & 0.090 & 0.071 & 0.026 & 74.2 & 13.4 & 79.1 & 13.4\\
 0.654 & 65.3 &  5.3 & 64.0 &  7.1 & -0.004 & 0.060 & 0.041 & 0.016 & 63.5 & 7.4 & 70.4 & 8.5\\
 0.604 & 68.3 &  5.2 & 61.7 &  7.7 & -0.157 & 0.056 & 0.051 & 0.018 & 52.0 & 9.3 & 69.4 & 8.1\\
 0.553 & 62.2 &  5.4 & 56.3 &  6.4 & -0.034 & 0.060 & 0.060 & 0.019 & 61.5 & 7.0 & 64.6 & 8.8\\
 0.502 & 64.6 &  4.4 & 54.8 &  6.7 & -0.022 & 0.053 & 0.070 & 0.022 & 63.9 & 5.9 & 64.2 & 7.6\\
 0.451 & 62.5 &  5.3 & 74.1 &  7.0 & -0.071 & 0.066 & 0.054 & 0.028 & 53.6 & 7.9 & 83.9 & 7.1\\
 0.401 & 63.9 &  5.0 & 61.8 &  7.6 & -0.040 & 0.048 & 0.089 & 0.018 & 61.8 & 7.4 & 75.3 & 10.6\\
 0.350 & 55.4 &  4.3 & 78.0 &  7.5 & -0.040 & 0.049 & 0.062 & 0.024 & 51.1 & 7.7 & 89.8 & 9.2\\
 0.299 & 60.1 &  4.5 & 76.8 &  7.0 & -0.178 & 0.046 & 0.053 & 0.023 & 37.4 & 8.9 & 86.8 & 6.5\\
\tablebreak
 0.249 & 62.5 &  3.9 & 66.5 &  6.8 & -0.021 & 0.039 & 0.080 & 0.029 & 60.1 & 5.2 & 79.6 & 7.2\\
 0.198 & 55.2 &  3.9 & 82.4 &  6.6 & -0.065 & 0.042 & 0.066 & 0.022 & 45.9 & 6.7 & 95.8 & 6.2\\
 0.147 & 43.3 &  4.7 & 101.2 & 6.0 & -0.004 & 0.048 & -0.022 & 0.035 & 42.5 & 7.9 & 95.8 & 6.4\\
 0.096 & 38.8 &  5.1 & 110.3 & 6.3 & -0.009 & 0.044 & 0.032 & 0.032 & 41.0 & 8.3 & 118.9 & 7.1\\
 0.046 & 37.2 &  5.0 & 120.6 & 7.0 & -0.028 & 0.038 & 0.118 & 0.034 & 31.3 & 9.1 & 115.6 & 9.6\\
 -0.005 & -3.7 &  4.7 & 132.0 & 6.9 & -0.051 & 0.037 & 0.139 & 0.034 & -15.2 & 10.7 & 176.8 & 11.5\\
  -0.056 & -24.5 &   5.4 & 130.0 &   7.0 & -0.006 &  0.037 &  0.010 &  0.034 & -25.7 & 10.1 & 133.2& 9.7 \\
  -0.106 & -31.6 &   4.7 & 103.0 &   6.0 &  0.018 &  0.039 &  0.028 &  0.030 & -28.1 & 7.5 & 110.2 & 8.4 \\
  -0.157 & -39.2 &   4.9 &  92.4 &   6.8 &  0.026 &  0.042 &  0.076 &  0.028 & -35.9 & 8.0 & 109.7 & 8.5 \\
  -0.208 & -52.1 &   4.8 &  87.6 &   6.5 &  0.033 &  0.041 &  0.033 &  0.031 & -47.5 & 7.1 & 94.8 & 6.4 \\
  -0.259 & -63.7 &   4.9 &  79.8 &   7.0 &  0.012 &  0.042 &  0.062 &  0.030 & -62.1 & 6.4 & 91.9 & 7.9 \\
  -0.309 & -64.2 &   5.0 &  67.6 &   6.5 &  0.090 &  0.039 &  0.068 &  0.022 & -56.6 & 6.3 & 78.9 & 9.0 \\
  -0.360 & -65.6 &   5.8 &  68.0 &  10.4 &  0.089 &  0.049 &  0.122 &  0.030 & -56.0 & 11.5 & 88.3 & 12.4 \\
  -0.411 & -55.7 &   5.5 &  65.1 &   7.8 &  0.017 &  0.043 &  0.102 &  0.022 & -54.3 & 7.6 & 81.5 & 10.8 \\
  -0.461 & -59.9 &   4.9 &  58.8 &   6.3 &  0.071 &  0.055 &  0.067 &  0.025 & -53.7 & 6.0 & 68.4 & 7.4 \\
  -0.512 & -59.0 &   5.4 &  50.7 &   6.4 &  0.028 &  0.046 &  0.061 &  0.018 & -58.4 & 5.8 & 58.3 & 9.8 \\
  -0.563 & -53.9 &   6.5 &  72.4 &   8.5 &  0.097 &  0.061 &  0.058 &  0.025 & -40.3 & 9.6 & 82.7 & 10.4 \\
  -0.614 & -54.2 &   7.4 &  67.2 &  12.0 &  0.036 &  0.064 &  0.127 &  0.040 & -51.1 & 11.2 & 88.0 & 14.8 \\
  -0.664 & -49.5 &   6.0 &  49.3 &   7.6 &  0.028 &  0.057 &  0.077 &  0.030 & -48.4 & 9.3 & 58.5 & 12.2 \\
  -0.715 & -59.2 &   7.1 &  61.0 &   9.4 &  0.175 &  0.061 &  0.069 &  0.028 & -42.6 & 14.3 & 71.3 & 14.1 \\
  -0.816 & -50.8 &   6.8 &  55.6 &   9.5 &  0.042 &  0.055 &  0.113 &  0.029 & -49.6 & 10.5 & 71.0 & 13.8 \\
  -0.918 & -43.3 &   6.9 &  58.9 &  10.1 &  0.050 &  0.080 &  0.101 &  0.031 & -36.2 & 10.3 & 73.5 & 15.2 \\
  -1.019 & -50.7 &   5.6 &  60.6 &   6.8 & -0.004 &  0.065 &  0.076 &  0.041 & -50.7 & 8.2 & 72.0 & 9.1 \\
  -1.121 & -56.2 &   6.7 &  53.3 &   8.1 &  0.080 &  0.073 &  0.066 &  0.027 & -49.7 & 14.2 & 61.9 & 12.6 \\
\tablebreak
  -1.252 & -58.6 &   7.3 &  60.2 &   8.5 &  0.092 &  0.072 &  0.066 &  0.022 & -47.2 & 10.8 & 69.9 & 12.4 \\
  -1.425 & -55.7 &   9.5 &  88.1 &  18.1 &  0.035 &  0.077 &  0.182 &  0.061 & -48.3 & 17.0 & 127.4 & 21.6 \\
  -1.658 & -64.8 &   6.4 &  51.2 &   7.4 &  0.037 &  0.058 &  0.056 &  0.019 & -60.5 & 9.0 & 58.2 & 9.4 \\
  -1.932 & -56.9 &   8.1 &  58.0 &   8.1 &  0.009 &  0.094 &  0.076 &  0.041 & -54.9 & 9.8 & 68.8 & 9.9 \\
  -2.257 & -75.5 &   9.5 &  67.2 &  13.5 &  0.164 &  0.096 &  0.084 &  0.041 & -43.6 & 17.1 & 81.1 & 20.4 \\
  -2.642 & -57.6 &  10.3 &  62.1 &  11.9 &  0.037 &  0.105 &  0.068 &  0.036 & -47.4 & 15.9 & 72.5 & 16.9 \\
  -3.078 & -56.0 &   9.9 &  53.8 &  11.1 &  0.268 &  0.106 &  0.088 &  0.043 & -28.7 & 17.7 & 65.4 & 15.9 \\
  -3.555 & -41.6 &  27.4 &  75.6 &  50.9 &  0.070 &  0.283 &  0.404 &  0.302 & -28.8 & 34.7 & 150.3 & 30.9 \\
  -4.194 &  -9.2 &  15.6 &  56.4 &  14.9 & -0.144 &  0.105 &  0.021 &  0.037 & -22.5 & 34.1 & 59.3 & 20.1 \\
  -4.569 & -39.7 &  10.8 &  45.3 &   9.2 &  0.240 &  0.136 &  0.042 &  0.036 & -19.7 & 17.4 & 49.9 & 13.2 \\
\enddata
\end{deluxetable}

\clearpage

\figcaption[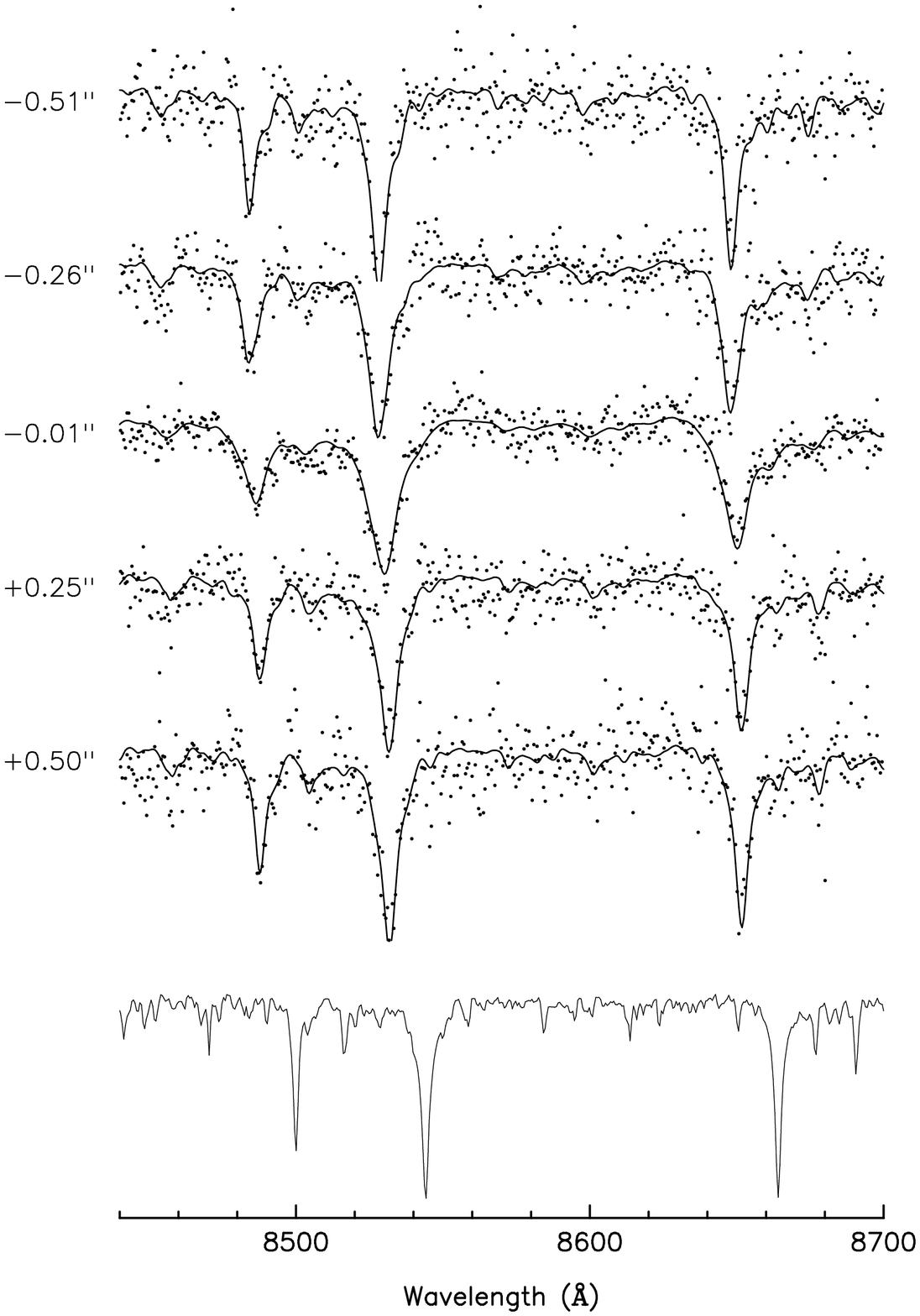]{\label{fig_spectra}}
Top: STIS spectra of M32 at five positions along the slit.
Solid curves are convolutions of the MPL-derived broadening functions 
$\hat{N}(V)$ (Fig. \ref{fig_broad}) with the stellar template.
Bottom: Spectrum of HR7615, a K0III giant, the template spectrum.
The vertical scale of the template spectrum is compressed with respect to that of the M32 spectra.

\figcaption[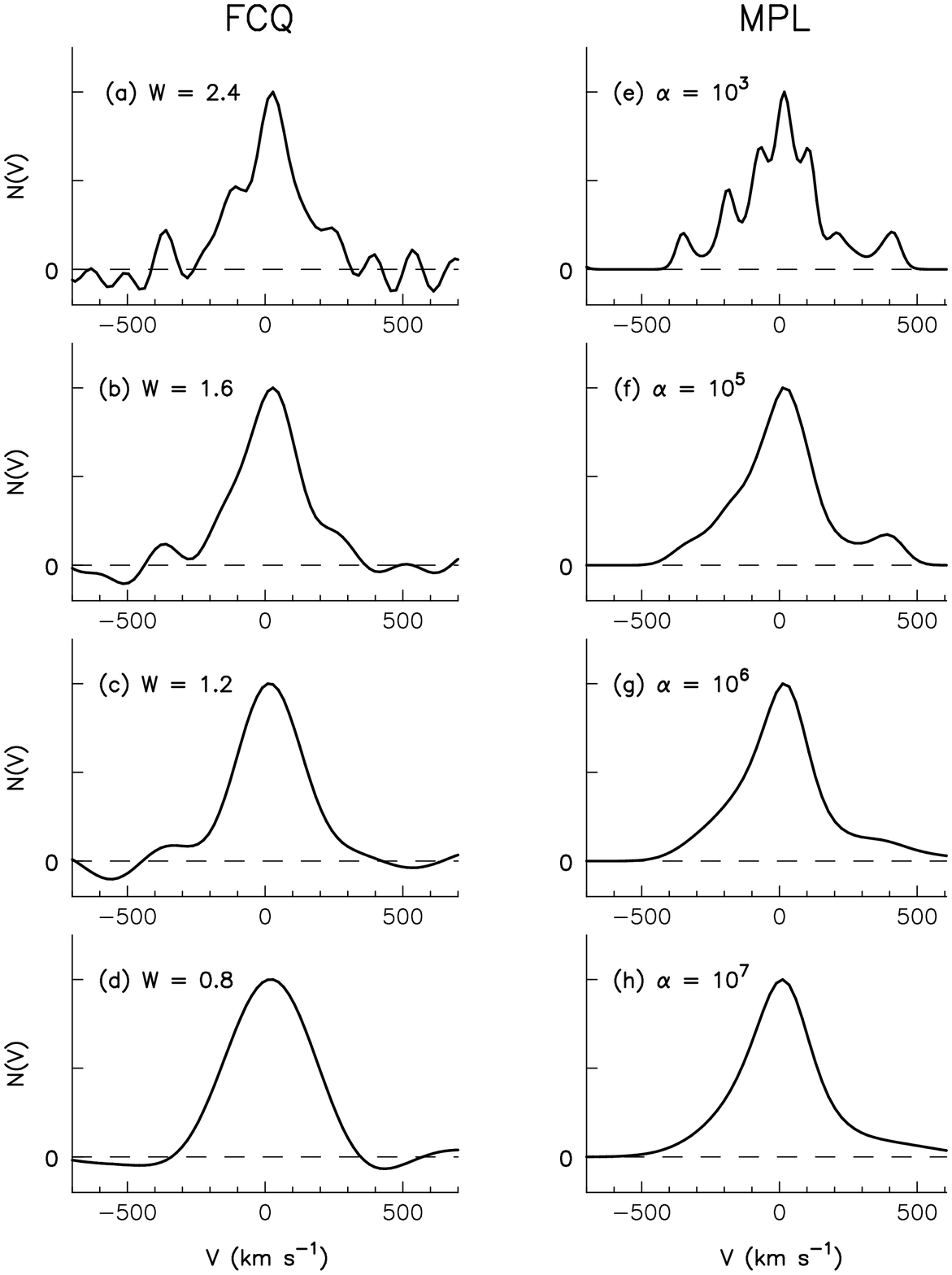]{\label{fig_compare}}
Broadening functions recovered from the central spectrum of M32 using the two spectral deconvolution algorithms discussed in the text.
The degree of smoothing increases downward.
Left column: FCQ. (a) $W=2.4$; (b) $W=1.6$; (c) $W=1.2$; (d) $W=0.8$.
Right column: MPL. (e) $\alpha=10^3$; (f) $\alpha=10^5$; (g) $\alpha=10^7$; (h) $\alpha=10^9$.
The MPL estimates tend toward a Gaussian for large $\alpha$ while the FCQ estimates become increasingly distorted as the smoothing is increased.
This is the source of the greater bias in the FCQ estimates (although in practice smoothing parameters as small as $W=0.8$ would never be used).

\figcaption[fig_h4.ps]{\label{fig_h4}}
The mean value of $h_4$ computed by the FCQ algorithm between $0.5''$ and $2.0''$, as a function of smoothing parameter $W$.
Large values of $W$ correspond to small degrees of smoothing and hence to less biased estimates.
These $h_4$ values are based on data that were heavily binned in radius in order to increase the S/N as much as possible.

\figcaption[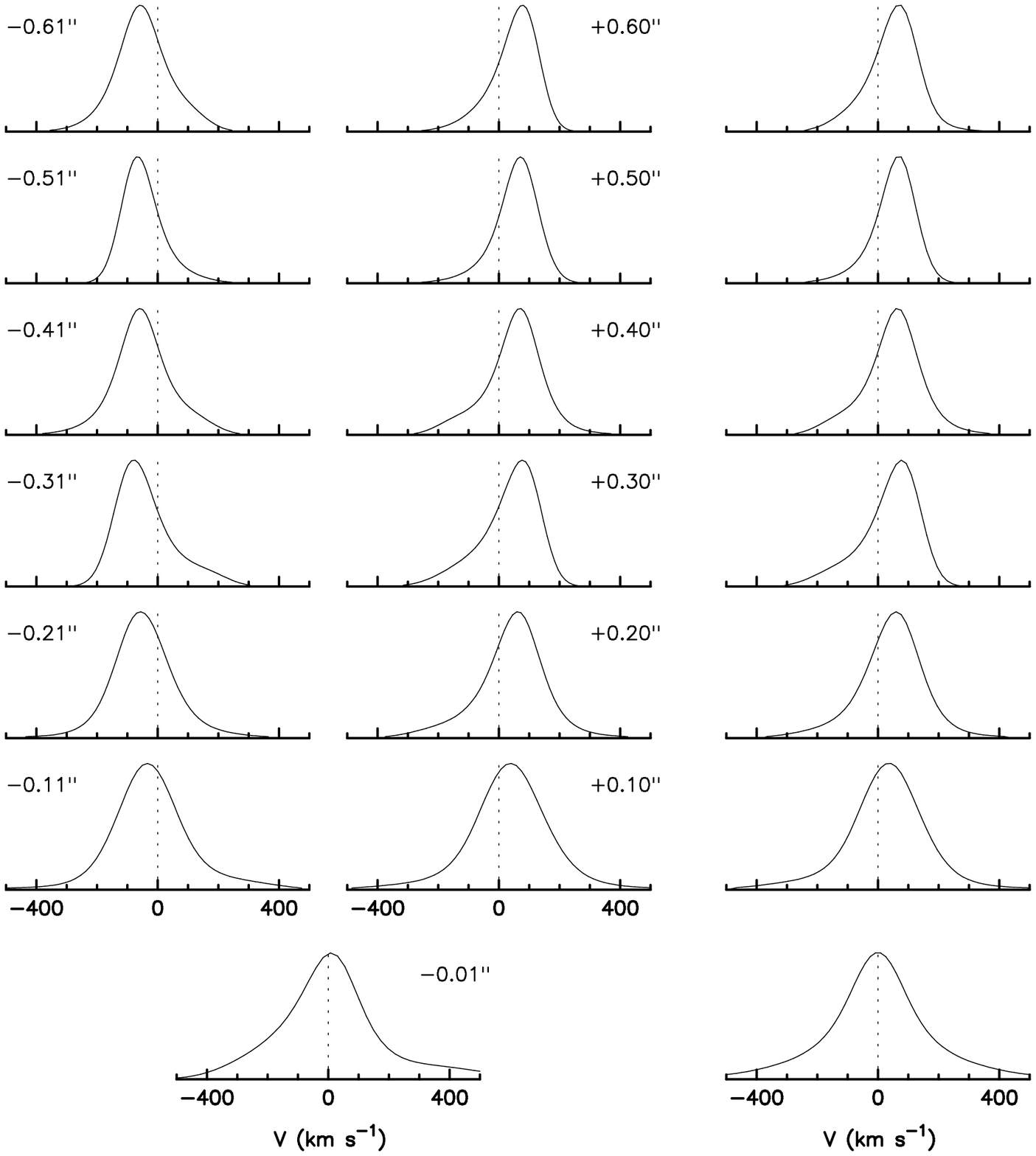]{\label{fig_broad}}
Line-of-sight velocity distributions derived from the STIS M32 spectra using the MPL deconvolution algorithm.
Note the sudden increase in the width of the broadening functions inside of $\sim 0.2''$.
The LOSVDs are roughly antisymmetric about the center of M32, as expected for a relaxed system;
the right column shows $\hat{N}(V)$ averaged over the left and right sides, ${1\over 2}[\hat{N}(V,R) + \hat{N}(-V,-R)]$.
The central LOSVD exhibits strong non-Gaussian wings, a likely consequence of high-velocity stars near the central black hole.
The broadening functions at larger radii exhibit asymmetries suggestive of a second kinematic subcomponent which rotates with a velocity closer to the systemic velocity of M32.

\figcaption[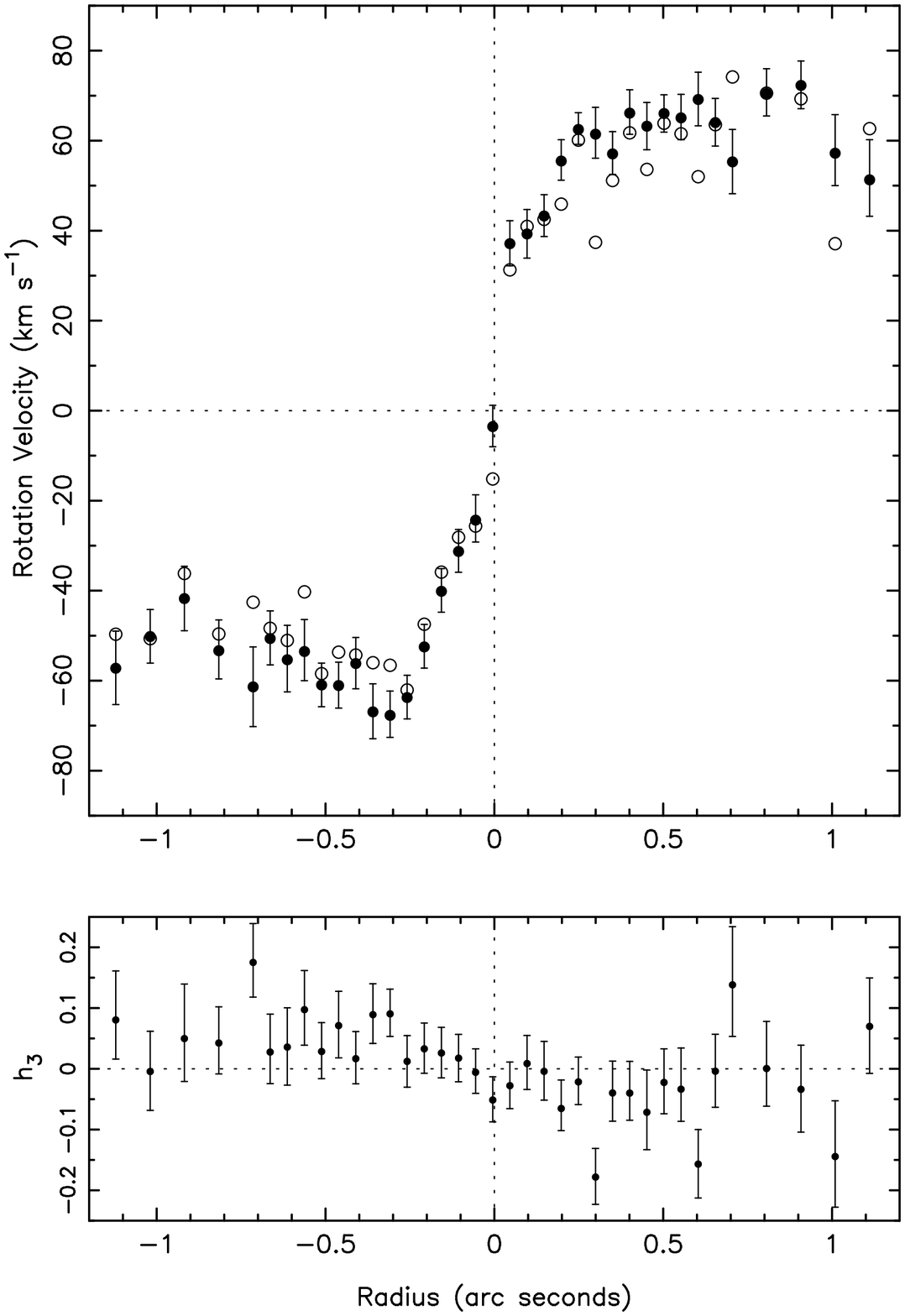]{\label{fig_rotate}}
STIS rotation curve for M32, derived from LOSVDs obtained using the MPL spectral deconvolution algorithm.
Upper panel: filled circles: $V_0$, the parameter that measures the velocity shift of the Gaussian function that multiplies the Gauss-Hermite series.
Open circles: $V_0 + \sqrt{3}\sigma_0 h_3$, an estimate of the true mean line-of-sight velocity.
Lower panel: the Gauss-Hermite parameter $h_3$ that measures asymmetries in the LOSVDs.
The mean velocity is smaller than $|V_0|$ due to the nonzero value of $h_3$, which in turn reflects asymmetries in the LOSVD's (Fig. \ref{fig_broad}).

\figcaption[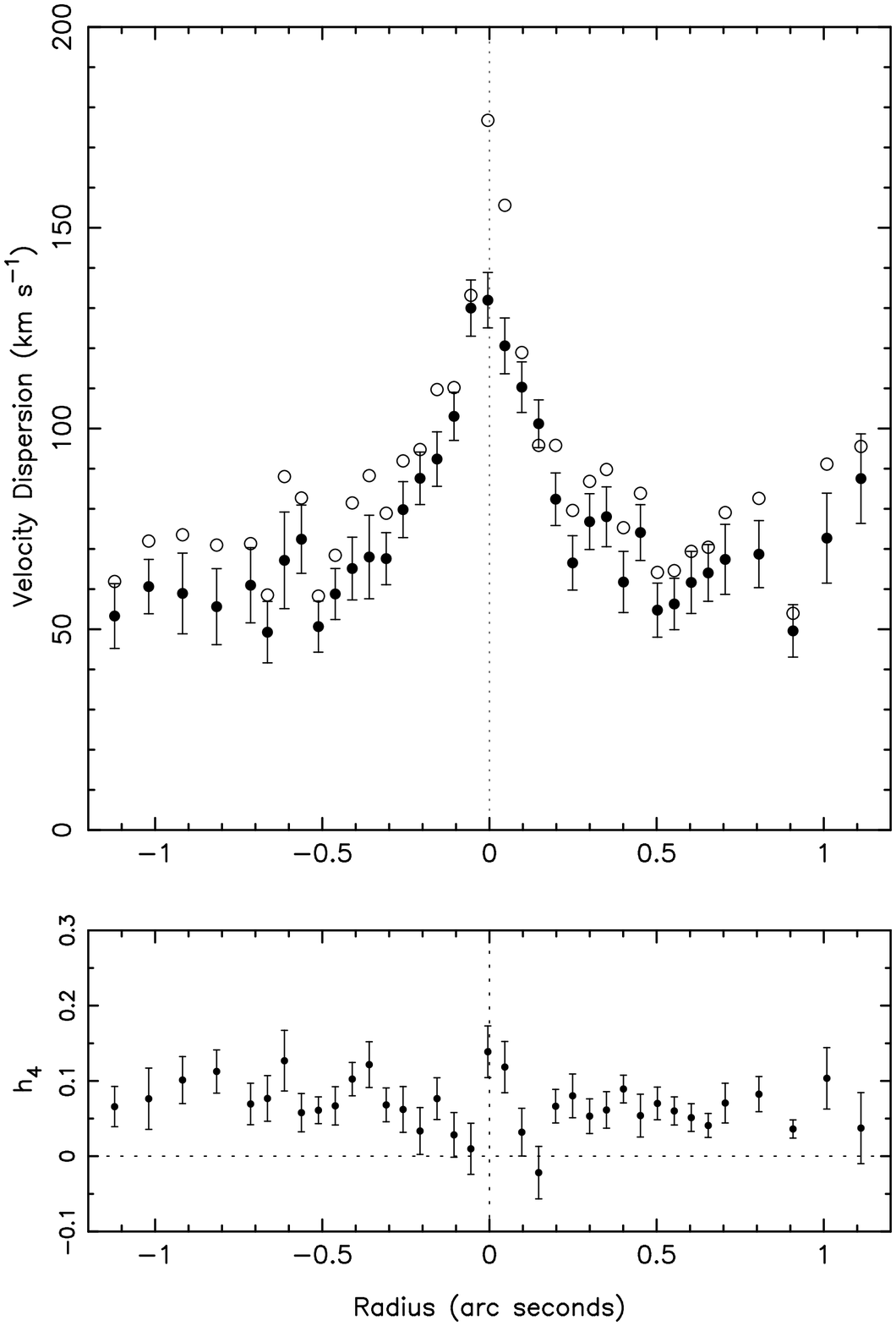]{\label{fig_dispersion}}
STIS velocity dispersion profile for M32, derived from LOSVDs obtained using the MPL spectral deconvolution algorithm.
Upper panel: filled circles: $\sigma_0$, the parameter that measures the dispersion of the Gaussian function that multiplies the Gauss-Hermite series.
Open circles: $\sigma_0(1+\sqrt{h_4})$, an estimate of the true rms line-of-sight velocity.
Lower panel: the Gauss-Hermite parameter $h_4$ that measures the amplitude of symmetric non-Gaussian distortions in the LOSVD.
The velocity dispersion is generally greater than $\sigma_0$ due to the nonzero values of $h_4$.
This difference is substantial in the inner $\sim 0.2''$ due to the strongly non-Gaussian wings of the central LOSVDs (Fig. \ref{fig_broad}).

\figcaption[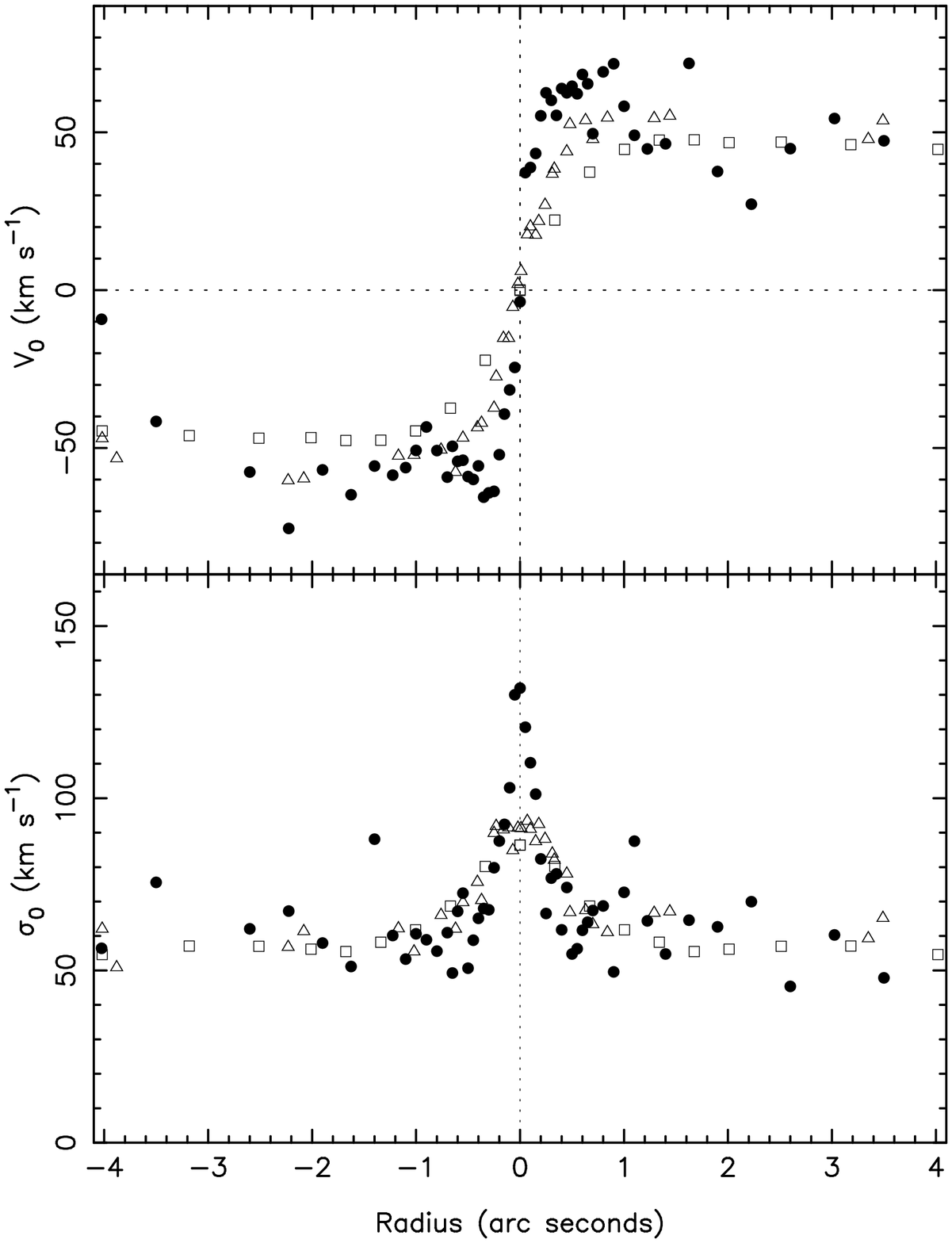]{\label{fig_gb}}
Comparison of $V_0$ and $\sigma_0$ derived from the M32 STIS data (filled circles) with earlier ground-based determinations.
Squares: WHT measurements from van der Marel et al. (1994a). 
Triangles: CFHT measurements from Bender et al. (1996).

\figcaption[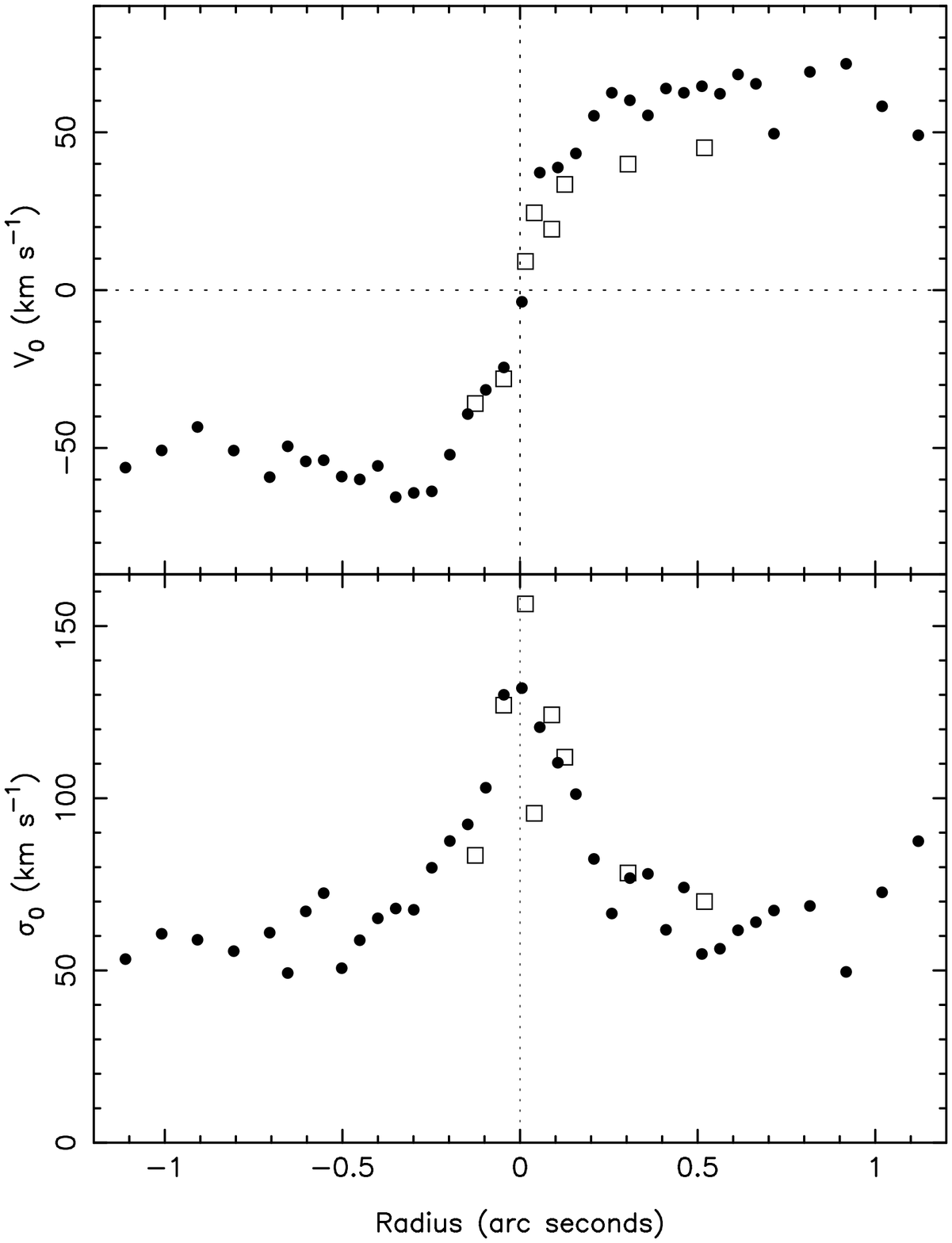]{\label{fig_fos}}
Comparison of $V_0$ and $\sigma_0$ derived from the M32 STIS data (filled circles) with FOS data of van der Marel et al. (1997) (open squares).

\figcaption[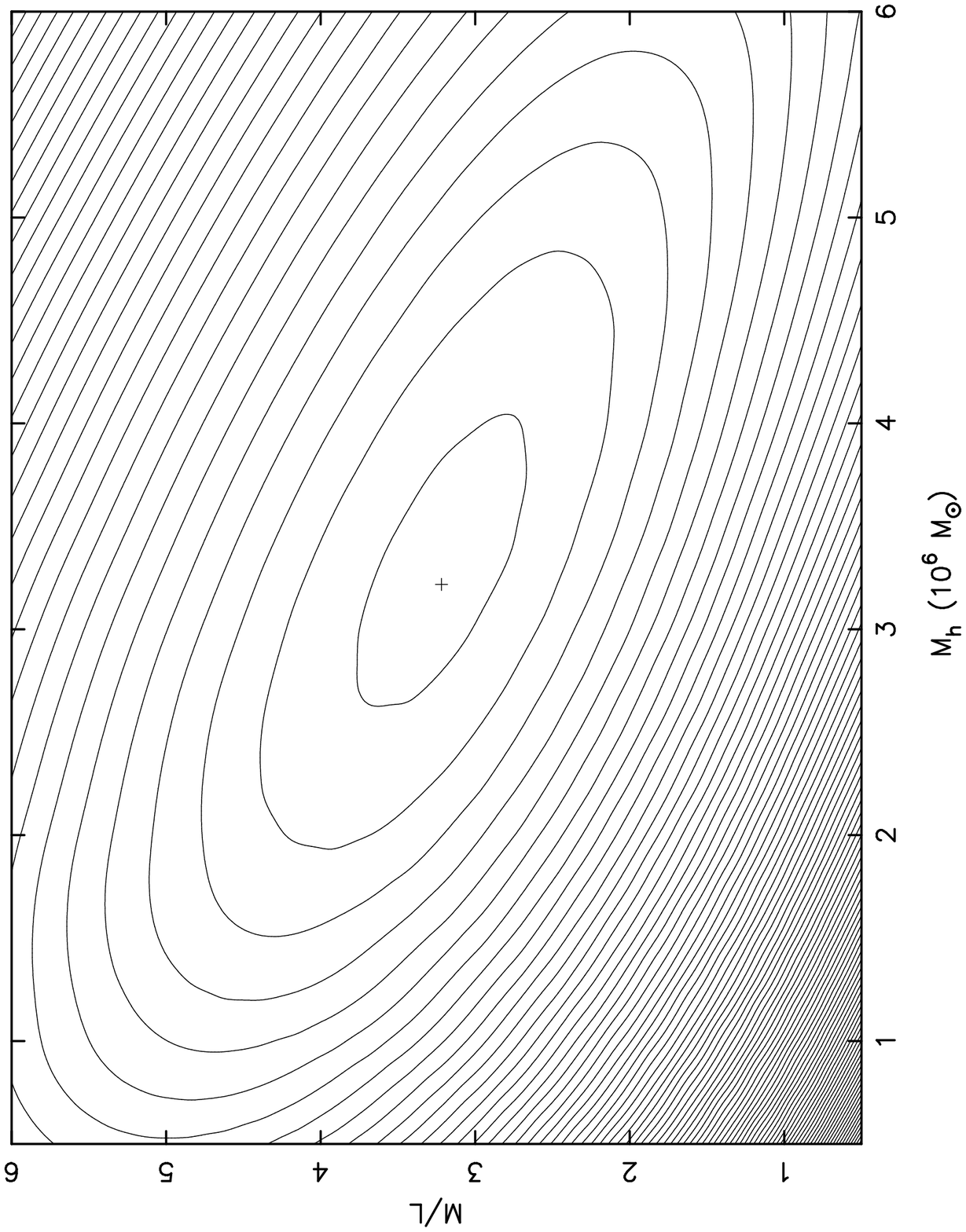]{\label{fig_contour}}
Reduced $\chi^2$ contours describing the fit of the axisymmetric models described in the text to the observed, mean square line-of-sight velocity at points within the inner arc second of M32.
The contour spacing is 0.5 and the innermost contour is at 0.75.
The plus symbol marks the best-fit model.

\figcaption[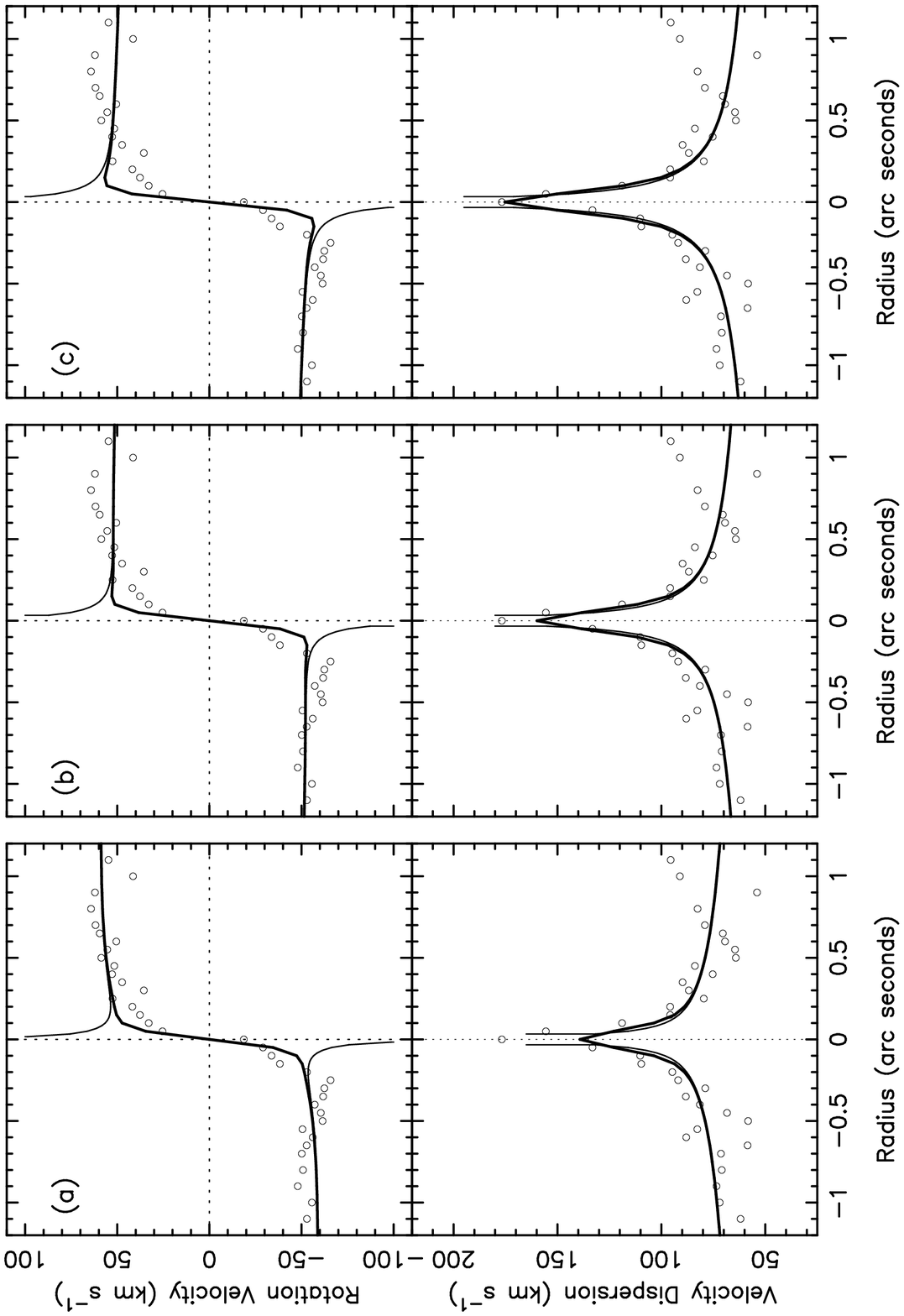]{\label{fig_three}}
Predicted kinematical profiles for three axisymmetric models with different black hole masses.
(a) $M_h=2.0\times 10^6\Msolar$;
(b) $M_h=3.0\times 10^6\Msolar$;
(c) $M_h=4.0\times 10^6\Msolar$.
Thin curves show the models as observed with infinite resolution; heavy curves are the models after convolution with the STIS PSF; open circles are the data points.
For each $M_h$, the mass to light ratio $M/L$ and rotational parameter $k$ have been adjusted to optimize the fit.

\figcaption[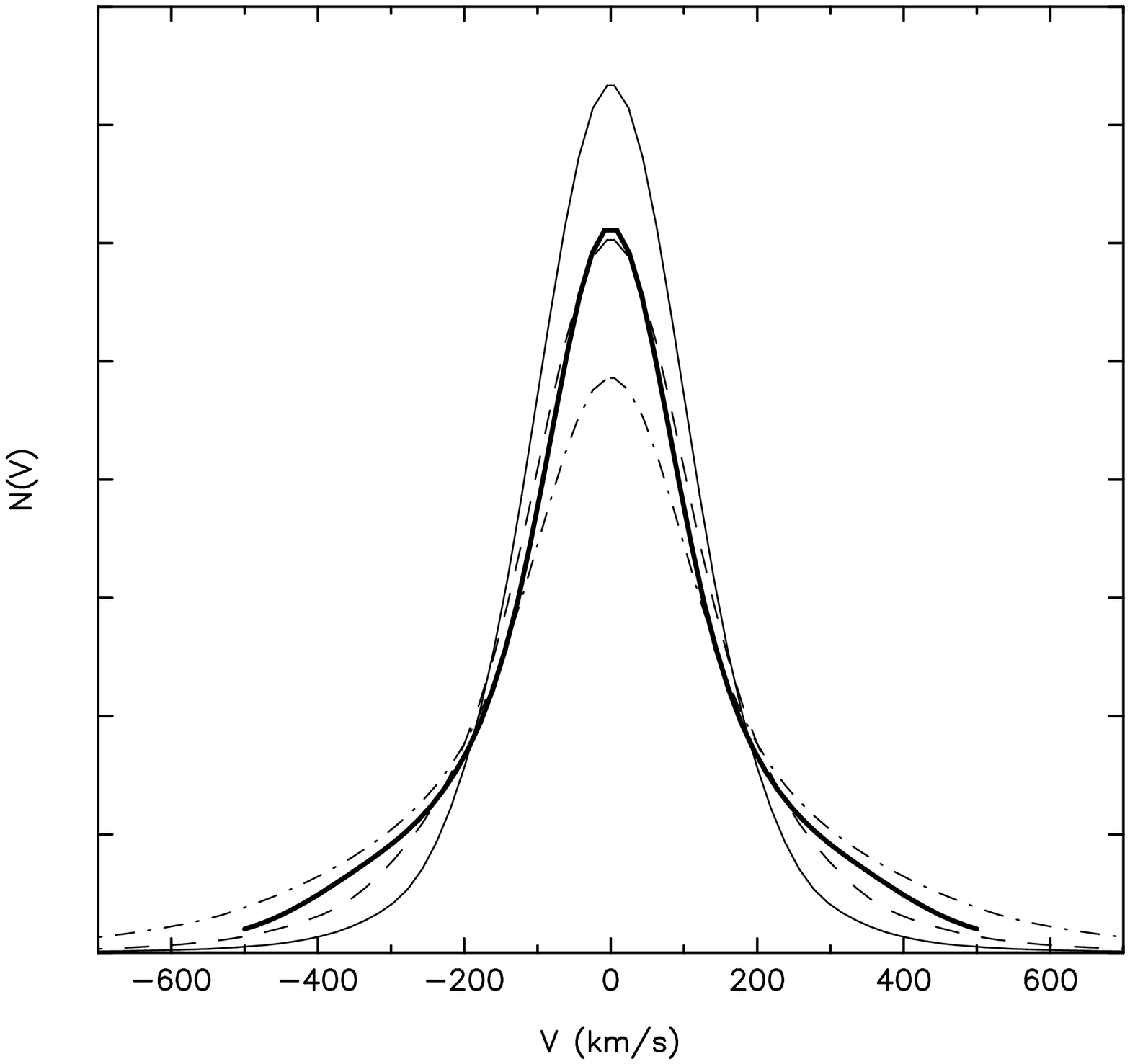]{\label{fig_center}}
The central M32 LOSVD (heavy line), symmetrized about $V=0$, 
compared to the LOSVD's predicted by spherical nonrotating models with three black holes masses.
Thin line: $M_h=2.5\times 10^6\Msolar$;
dashed line: $M_h=5.0\times 10^6\Msolar$;
thin line: $M_h=10.\times 10^6\Msolar$.

\noindent Fig. A1 ---

Estimated Gauss-Hermite parameters $\hat\sigma$ and $\hat{h_4}$, derived by fitting the $N(V)$ of equation (A6) to the assumed form (A5), with $j_{max}=4$.
Filled circles indicate the input values of $\sigma_0$ and $h_4$; these values are recovered only when the input $N(V)$ has $h_6=0$.

\noindent Fig. B1 ---

Recovery of $\sigma_0$ via FCQ.

\noindent Fig. B2 ---

Recovery of $h_4$ via FCQ.
Circles: $\sigma_0=40$ \kms.
Triangles: $\sigma_0=60$ \kms.
Diamonds: $\sigma_0=100$ \kms.
There is a significant negative bias in the recovered values of $h_4$ when the velocity dispersion is less than about 100 \kms.

\noindent Fig. B3 ---

Mean estimates of $N(V)$ averaged over 100 random realizations of the observed spectrum, for S/N $=\{5,20,100\}$.
The input $N(V)$ (Equation B2) is shown by the heavy curves.

\noindent Fig. B4 ---

MISE and ISB of estimates of $N(V)$ obtained from the two deconvolution algorithms.
The input LOSVD was a Lorentzian (Equation B2) 
with $\sigma_0=108 $\kms\ and $h_4=0.15$.
Both MISE and ISB have been normalized as described in the text.
Solid lines: MPL algorithm.
Open circles: FCQ algorithm, using a fixed smoothing parameter $W=1.3$.
Filled circles: FCQ algorithm, using the value $W_{opt}$ that minimizes the MISE of the estimated $N(V)$.

\noindent Fig. B5 ---

MSE and bias in estimates of $h_4$ obtained from the two deconvolution algorithms.
The input $N(V)$ was a Lorentzian (Equation B2)
with $\sigma_0=108 $\kms\ and $h_4=0.15$.
Solid lines: MPL algorithm.
Open circles: FCQ algorithm, with $W=1.2$.
Filled circles: FCQ algorithm, using the value $W_{opt}$ that minimizes the MISE of the estimate $\hat{h}_4$.
$W_{opt}$ is plotted vs. S/N in the bottom panel.

\clearpage

\setcounter{figure}{0}

\begin{figure}
\plotone{fig_spectra2.ps}
\caption{ }
\end{figure}

\begin{figure}
\plotone{fig_compare.ps}
\caption{ }
\end{figure}

\begin{figure}
\plotone{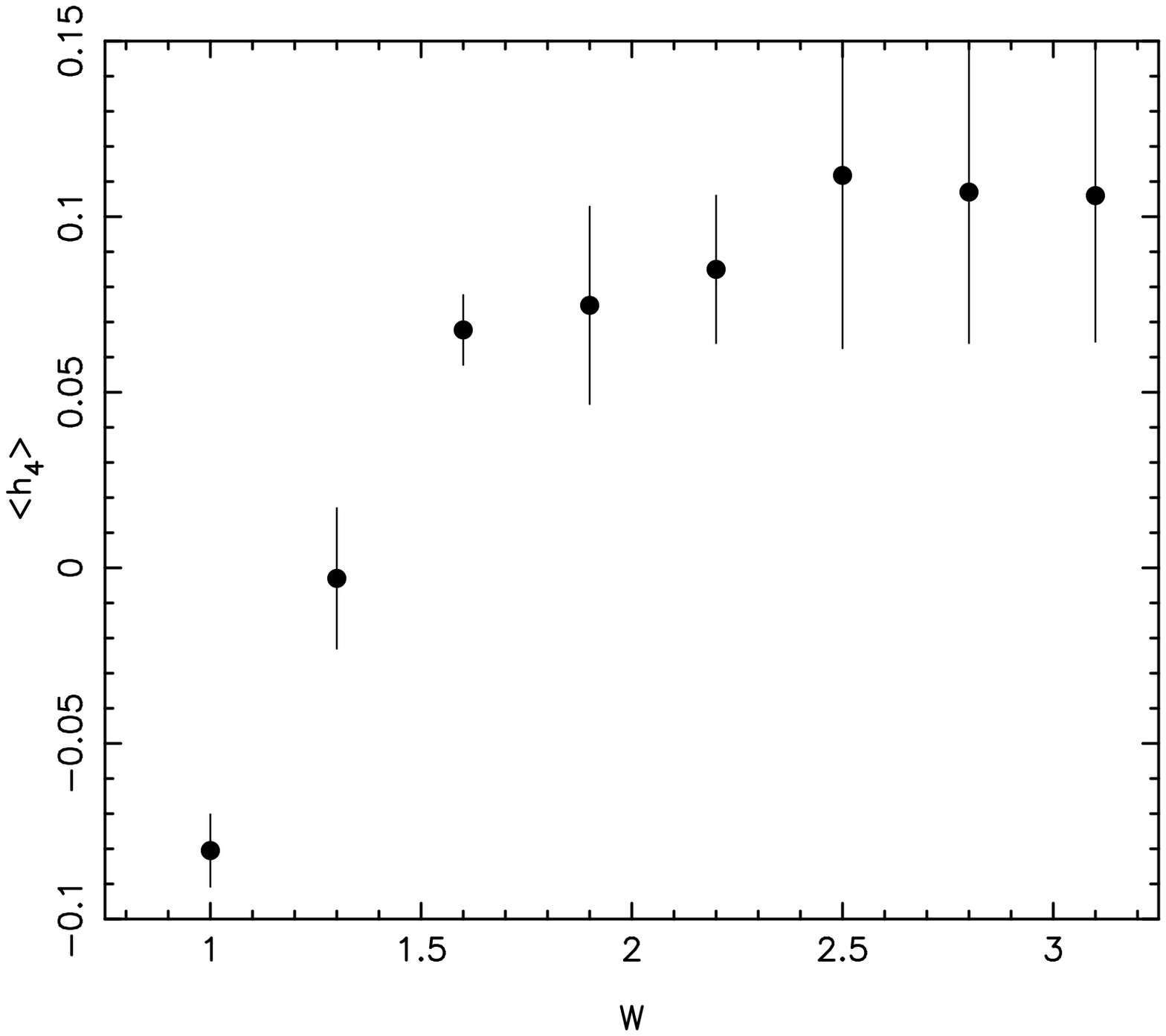}
\caption{ }
\end{figure}

\begin{figure}
\plotone{fig_broad2.ps}
\caption{ }
\end{figure}

\begin{figure}
\plotone{fig_rot3.ps}
\caption{ }
\end{figure}

\begin{figure}
\plotone{fig_disp2.ps}
\caption{ }
\end{figure}

\begin{figure}
\plotone{fig_gb2.ps}
\caption{ }
\end{figure}

\begin{figure}
\plotone{fig_fos2.ps}
\caption{ }
\end{figure}

\begin{figure}
\plotone{fig_cont1.ps}
\caption{ }
\end{figure}

\begin{figure}
\plotone{fig_three.ps}
\caption{ }
\end{figure}

\begin{figure}
\plotone{fig_center.ps}
\caption{ }
\end{figure}


\begin{figure}
\plotone{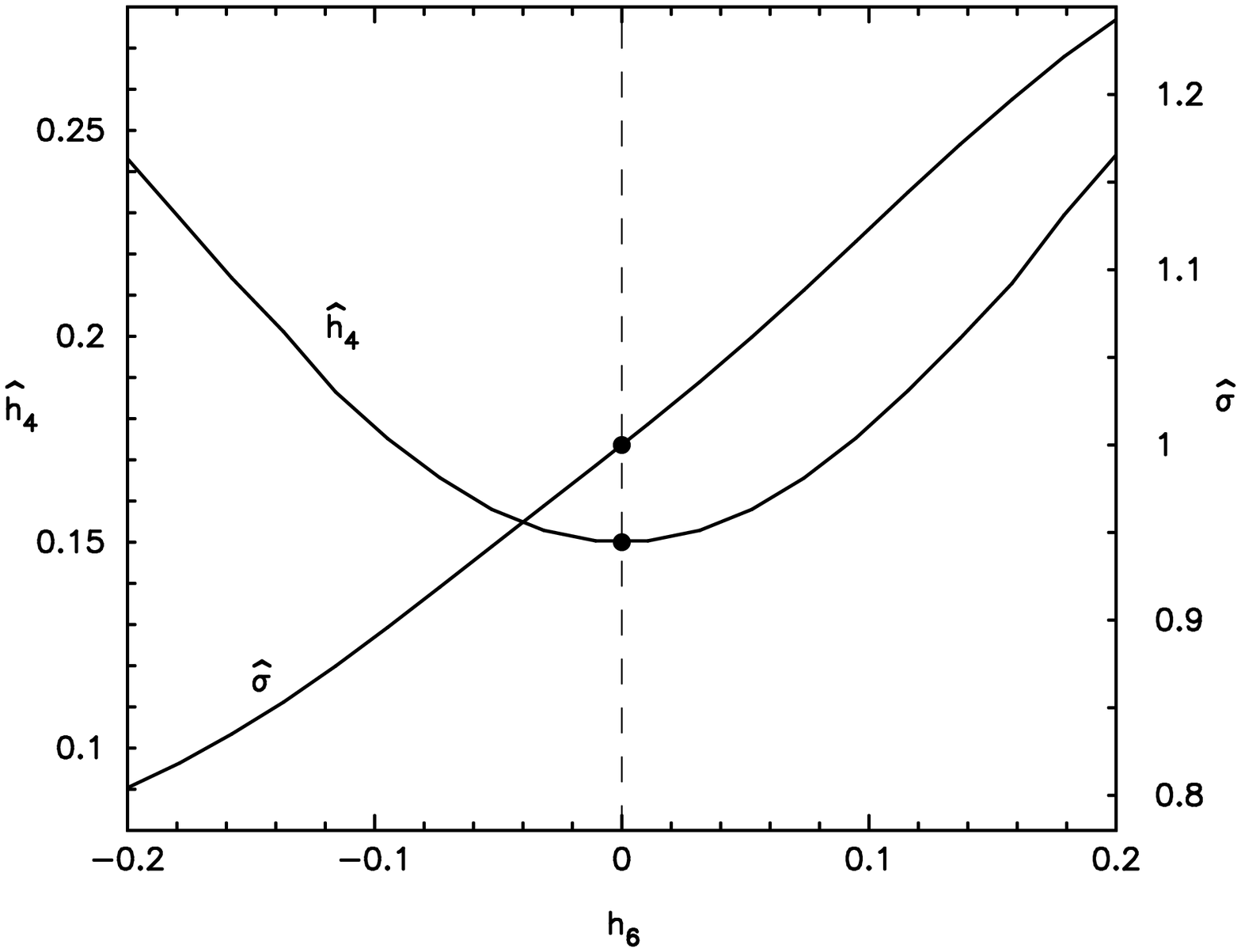}
Fig.~A1
\end{figure}

\begin{figure}
\plotone{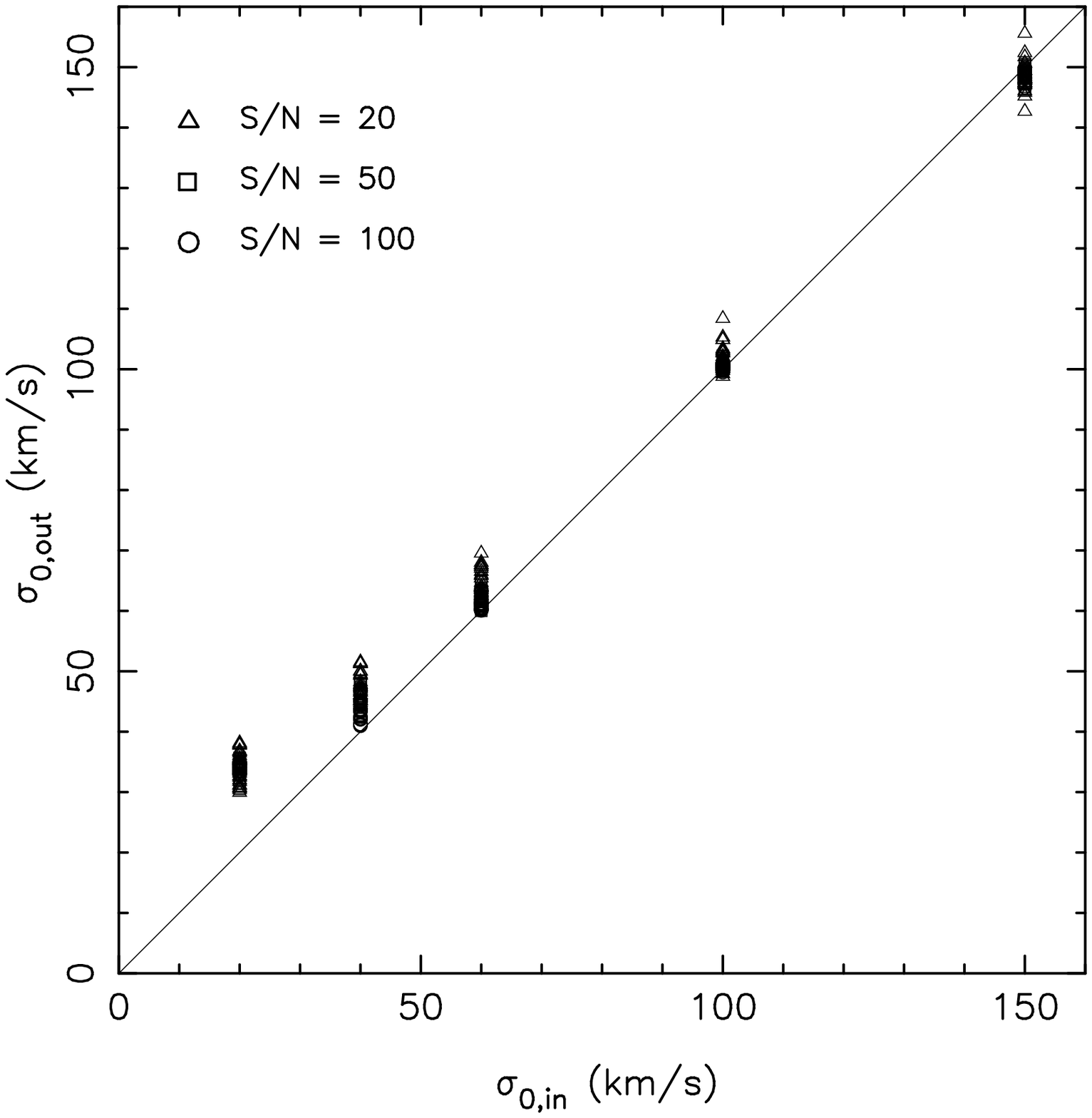}
Fig.~B1
\end{figure}

\begin{figure}
\plotone{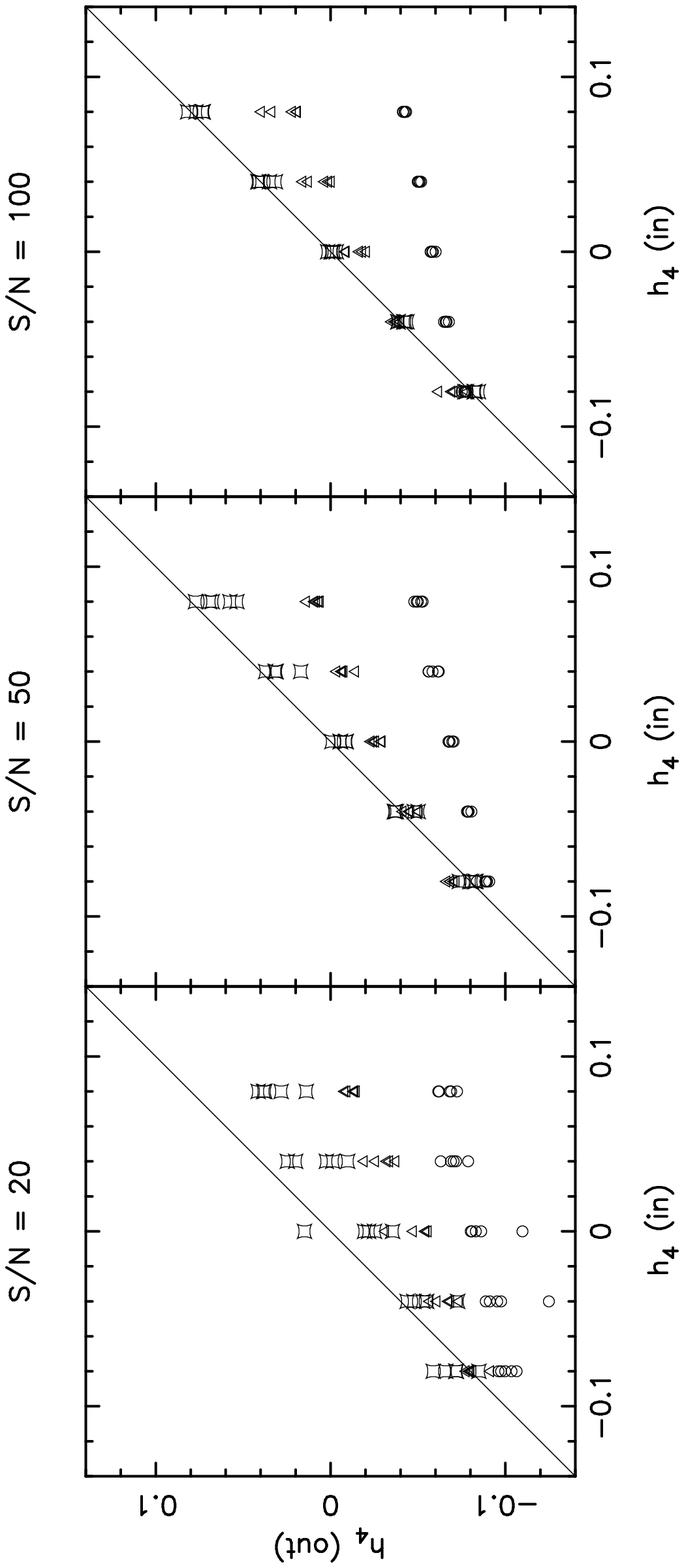}
Fig.~B2
\end{figure}

\begin{figure}
\plotone{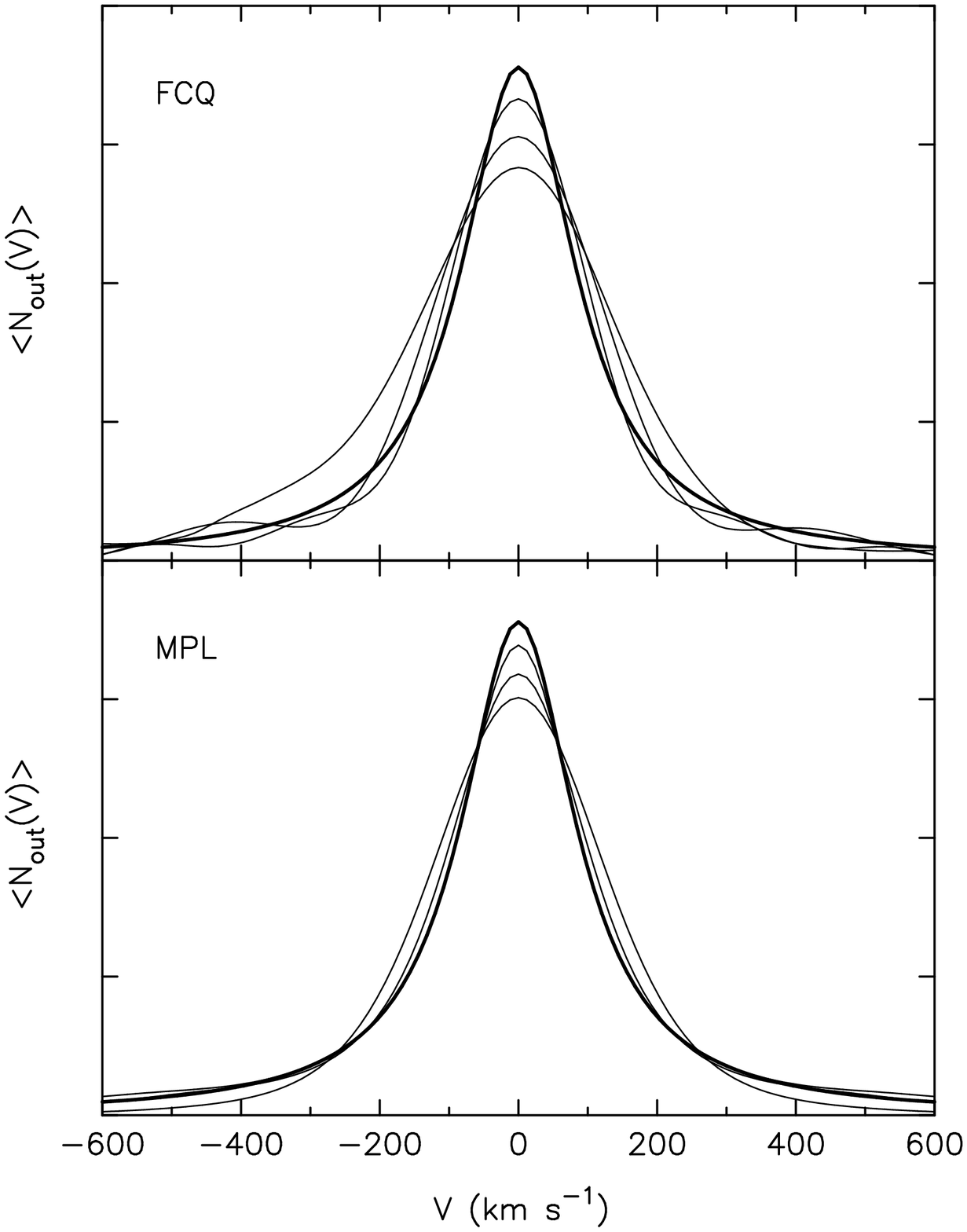}
Fig.~B3
\end{figure}

\begin{figure}
\plotone{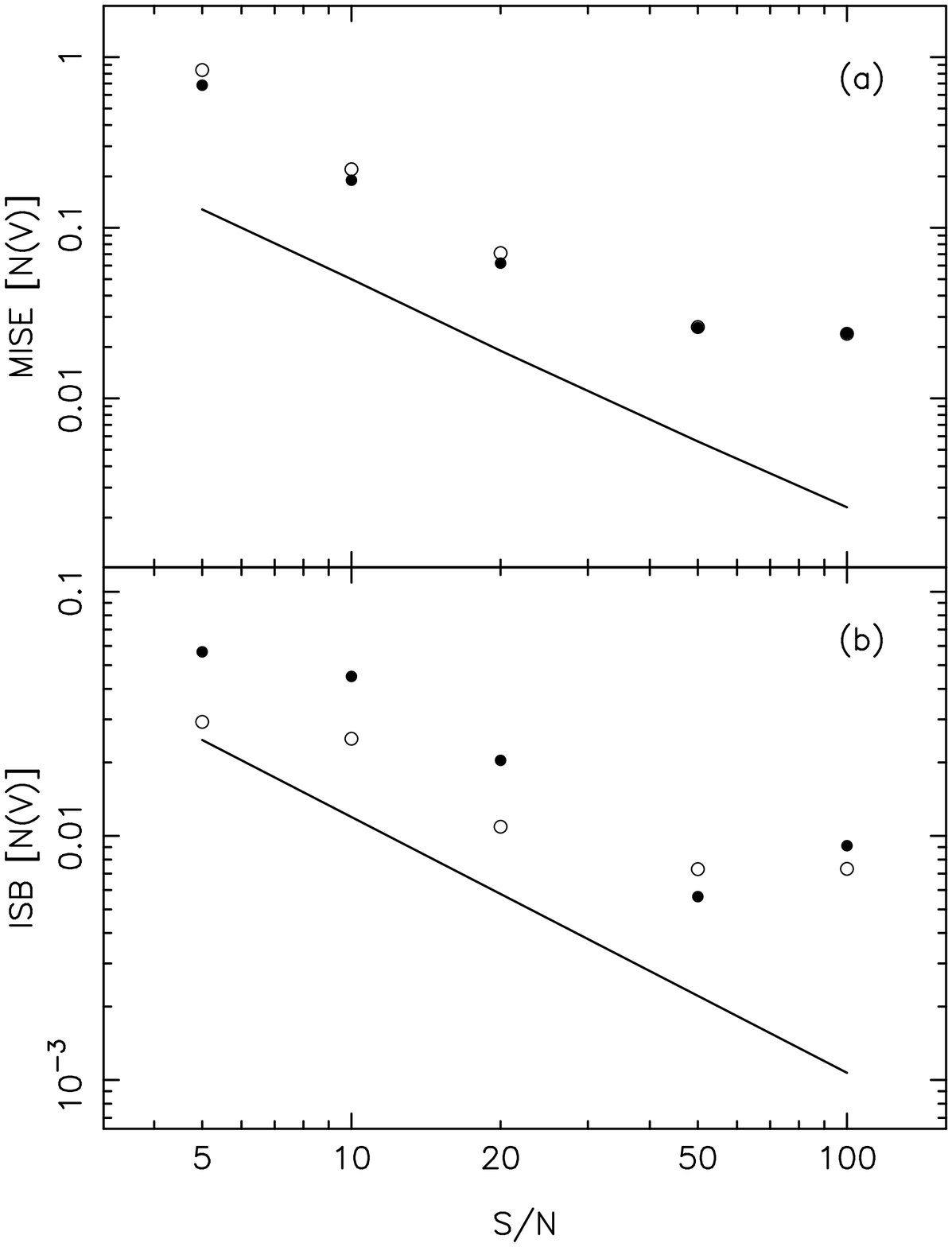}
Fig.~B4
\end{figure}

\begin{figure}
\plotone{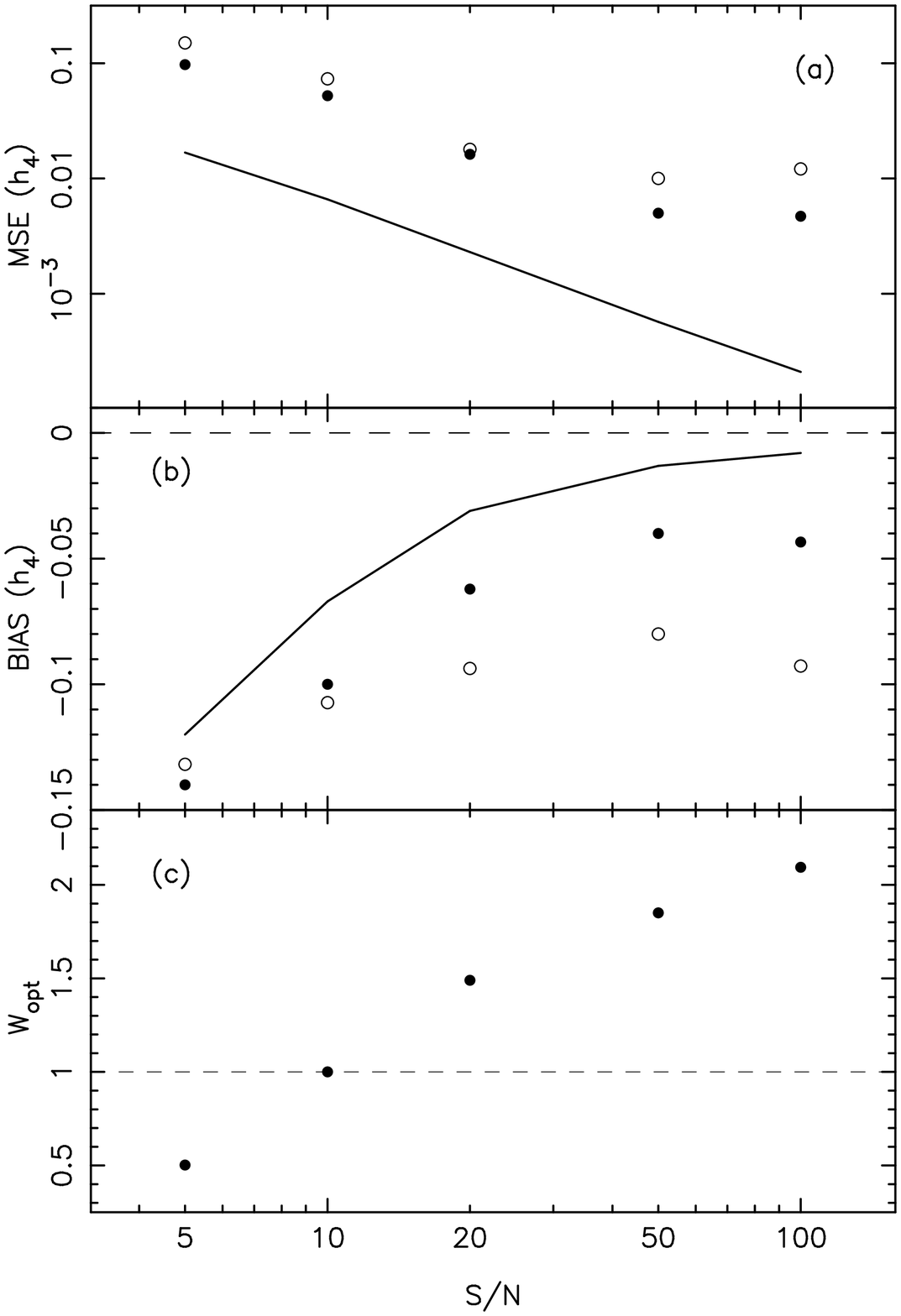}
Fig.~B5
\end{figure}

\clearpage


\clearpage
\begin{thebibliography}{}

\bibitem[Bahcall \& Wolf 1976]{baw76} Bahcall, J. \& Wolf, S. 1976,
	ApJ, 209, 214
\bibitem[Bender 1990] {ben90} Bender, R. 1990, A\&A, 229, 441
\bibitem[Bender, Kormendy \& Dehnen 1996] {bkd96} Bender, R., Kormendy, J.
	\& Dehnen, W. 1996, ApJ, 464, L123
\bibitem[Bender, Paquet \& Nieto 1991]{bpn91} Bender, R., Paquet, A. \& Nieto, 		J.-L. 1991, \AA, 246, 349
\bibitem[Bender, Saglia \& Gerhard 1994] {bsg94} Bender, R., Saglia, R.
	\& Gerhard, O. 1994, MNRAS, 269, 785
\bibitem[Bower et al. 1998]{bow98} Bower, G. A. 1998, ApJ, 492, L111
\bibitem[Bower et al. 2000]{bow00} Bower, G. A. et al. 2000, in preparation
\bibitem[Carter \& Jenkins 1993]{caj93} Carter, D. \& Jenkins, C. R. 1993,
	MNRAS, 263, 1049
\bibitem[Dehnen 1995] {deh95} Dehnen, W. 1995, MNRAS, 274, 919
\bibitem[Dressler \& Richstone 1988]{drr88} Dressler, A. \& Richstone,
	D. O. 1988, ApJ, 324, 701
\bibitem[Eddington 1916]{edd16} Eddington, A. S. 1916, MNRAS, 76, 572
\bibitem[Ferrarese \& Ford 1999] {fef99} Ferrarese, L. \& Ford, H. C. 1999,
	ApJ, 515, 583
\bibitem[Fillmore 1986]{fil86} Fillmore, J. A. 1986, AJ, 91, 1096
\bibitem[Gerhard 1993]{ger93} Gerhard, O. E. 1993, MNRAS, 265, 213
\bibitem[Goudfrooij, Baum \& Walsh 1997]{gbw97} Goudfrooij, P.,
	Baum, S. A. \& Walsh, J. R. 1997, in 
	The 1997 HST Calibration Workshop with a New Generation of Instruments,
	100.
\bibitem[Kimble et al. 1998]{kim98} Kimble, R. A. et al. 1998,
	ApJ L, 492, 283
\bibitem[Merritt 1987]{mer87} Merritt, D. 1987, ApJ, 313, 121
\bibitem[Merritt 1993]{mer93} Merritt, D. 1993, ApJ, 413, 79 
\bibitem[Merritt 1997]{mer97} Merritt, D. 1997, AJ, 114, 228
\bibitem[Merritt 1999]{mer99} Merritt, D. 1999, PASP, 111, 129
\bibitem[Myller-Lebedeff 1908]{myl08} Myller-Lebedeff, W. 1908, 
	Math. Ann. 64, 388
\bibitem[Peebles 1972]{pee72} Peebles, P. J. E. 1972, Gen. Rel. Grav.,
	3, 63
\bibitem[Qian et al. 1995]{qia95} Qian, E. E., de Zeeuw, P. T., van 
	der Marel, R. P. \& Hunter, C. 1995, MNRAS, 274, 602
\bibitem[Sargent et al. 1977]{sar77} Sargent, W. L. W., Schechter, P. L., 
	Boksenberg, A. \& Shortridge, K. 1977, ApJ, 212, 326
\bibitem[Satoh 1980]{sat80} Satoh, C. 1980, PASJ, 32, 41
\bibitem[Silverman 1982]{sil82} Silverman, B. W. 1982, Ann. Stat. 10, 795
\bibitem[Silverman 1986]{sil86} Silverman, B. W. 1986, 
	Density Estimation for Statistics and Data Analysis
	(Chapman \& Hall: London)
\bibitem[Thompson \& Tapia 1990]{tht90} Thompson, J. R. \& Tapia,
	R. A. 1990, Nonparametric Function Estimation,
	Modeling, and Simulation (SIAM, Philadelphia), 37
\bibitem[Tonry 1987]{ton87} Tonry, J. L. 1987, ApJ, 322, 632
\bibitem[van der Marel 1994]{vdm94} van der Marel, R. 1994,
	ApJ, 432, 91
\bibitem[van der Marel \& Franx 1993]{vdf93} van der Marel, R. \&
	Franx, M. 1993, ApJ, 407, 525
\bibitem[van der Marel et al. 1994a]{vdm94a} van der Marel, R. P., 
	Rix, H.-W., Carter, D., Franx, M., White, S. D. M. \& de Zeeuw, T.
	1994a, MNRAS, 268, 521
\bibitem[van der Marel et al. 1994b]{vdm94b} van der Marel, R. P., 
	Wyn-Evans, N., Rix, H.-W., White, S. D. M. \& de Zeeuw, T.
	1994b, MNRAS, 271, 99 
\bibitem[van der Marel et al. 1997]{vdm97} van der Marel, R. P., 
	de Zeeuw, P. T. \& Rix, H. W. 1997, ApJ, 488, 119
\bibitem[van der Marel et al. 1998]{vdm98} van der Marel, R. P., 
	Cretton, N., de Zeeuw, P. T. \& Rix, H. W. 1998, ApJ, 493, 613
\bibitem[Woodgate et al. 1998]{woo98} Woodgate, B. E. et al. 1998,
	PASP, 110, 1183
\end{thebibliography}
\end {document}